%% file: main_v1.tex
\documentclass[journal]{new-aiaa} 

\usepackage[utf8]{inputenc}
\usepackage{textcomp}
\usepackage{graphicx}
\usepackage{amsmath}
\usepackage[version=4]{mhchem}
\usepackage{siunitx}
\usepackage{longtable,tabularx}
\usepackage{amsmath,systeme}
\usepackage{cancel}
\usepackage{mathrsfs}
\usepackage{caption}
\usepackage{subcaption}
\usepackage{graphicx}
\usepackage{amsthm}
\usepackage{algpseudocode}
\usepackage{xcolor}
\usepackage{color,soul}
\usepackage{lscape}
\usepackage{algorithm}
\usepackage{comment}
\usepackage{tikz}
\usetikzlibrary{shapes.geometric, arrows.meta, positioning} 

\tikzstyle{startstop} = [ellipse, draw, fill=blue!20, text centered, minimum width=3cm]
\tikzstyle{process} = [rectangle, draw, fill=green!20, text centered, minimum width=4cm]
\tikzstyle{decision} = [diamond, draw, fill=orange!20, text centered, aspect=2]
\tikzstyle{arrow} = [thick, ->, >=stealth]

\graphicspath{{Figs/}}

\newtheorem{rem}{Remark}

\title{Optimal Bounded Thrust Powered Descent \\ with Analytical Ground-Collision Avoidance}

\author{Or Nataf \footnote{Graduate student, The Stephen B. Klein Faculty of Aerospace Engineering, Technion. E-mail: ornataf@campus.technion.ac.il} and Vitaly Shaferman \footnote{Associate Professor, The Stephen B. Klein Faculty of Aerospace Engineering, Technion. E-mail: vitalysh@technion.ac.il, (Corresponding Author).}}
\affil{Technion - Israel Institute of Technology, Haifa 3200003, Israel}

\input{command}

\renewcommand{\include}{\input}

\begin{document}
\maketitle

\begin{abstract}
The paper proposes a new approach to address the bounded-thrust powered-descent problem while ensuring ground-collision avoidance. A time-dependent polynomial approximation of the mass is employed to formulate a bounded linear-quadratic optimal-control problem that minimizes the thrust-acceleration control effort, terminal miss, and terminal velocity error. The resulting approximation is used to impose a hard constraint on the horizontal thrust profile while keeping the vertical thrust profile unconstrained. The key idea is a hierarchical separation of the thrust allocation, which enables analytical ground-collision avoidance under bounded thrust. Unlike existing bounded-thrust powered-descent approaches based on numerical optimization and trajectory-shaping constraints, the proposed method provides explicit analytical collision-avoidance conditions. Building on this formulation, the guidance law predicts the switching times between saturated and unsaturated arcs and shapes the thrust-acceleration profile to achieve a soft landing, even when the controller remains saturated over extended portions of the trajectory. Owing to its analytical nature, the guidance law is computationally efficient, and its continuous thrust profile facilitates real-time implementation. The proposed method was evaluated over a grid of perturbed initial conditions in realistic simulations, demonstrating accurate collision-free soft-landing performance. The results highlight the importance of combining saturation-aware guidance with ground-collision avoidance under bounded thrust.

\end{abstract}

\section{Introduction} \label{Introduction}
Space exploration has expanded significantly in recent years. In particular, the topic of powered descent has advanced rapidly and attracted growing academic interest. The powered descent maneuver 
employs the lander's rocket engines to guide it to a soft landing at the designated landing site (i.e., zero velocity at the target coordinates). The ongoing demand for improved and guaranteed performance, as well as simple algorithms suitable for real-time spacecraft applications, has led researchers to propose various approaches to address the main challenges of the problem, including fuel-consumption minimization, bounded-thrust control, ground-collision avoidance, approach-angle control, and other mission requirements and constraints.  

The Apollo program fostered substantial contributions to powered descent research. Among the guidance laws considered during the Apollo program, Apollo Powered Descent Guidance (APDG) \cite{APDG}, and E-Guidance \cite{Eguidance} have gained significant recognition.
APDG was ultimately selected for the Apollo program. However, E-guidance was later shown to be an optimal guidance law that minimizes the thrust-acceleration control effort in a constant gravitational field \cite{1997}.
Both APDG and E-guidance assume that the thrust-acceleration vector is a time-dependent polynomial function. E-guidance assumes a linear polynomial for the thrust-acceleration vector in which the coefficients are determined by the terminal position and velocity conditions. Similarly, APDG assumes that the thrust-acceleration vector is a quadratic polynomial, with coefficients determined by the terminal acceleration conditions, in addition to the position and velocity.
Augmented Apollo Powered Descent Guidance (AAPDG) \cite{lu2019augmented} is a family of tunable guidance laws that includes APDG and E-guidance as special cases and allows adjusting the guidance law to meet mission requirements using a single gain.

Although numerous guidance laws have been proposed for this problem, optimal-control-based solutions are particularly desirable because they can be derived to optimize a predefined objective. In particular, some optimal guidance laws employ linear-quadratic (LQ) optimal control formulations. Among them is the guidance law in \cite{1997}, which employed an LQ formulation with an integral cost on the thrust-acceleration control effort, the terminal time, and constrained terminal conditions (i.e., position and velocity).
The LQ optimal control framework was also used in \cite{Gutman}; however, unlike in \cite{1997}, the time horizon was fixed, and the terminal conditions were softly constrained. A known limitation of these guidance laws is that they do not account for thrust bounds.

Accounting for thrust bounds in the powered descent guidance problem is important to ensure solution feasibility. Unaccounted-for saturation poses a risk to the lander and may result in a target miss and a ground collision.
Engine capacity limits impose an upper bound on the thrust vector magnitude, and a lower bound on the thrust magnitude is sometimes required because the engine cannot be fully throttled off. 
The explicit minimum-propellant, bounded-thrust optimal control problem was addressed in \cite{PropellantOptimal}. The derived guidance law had a bang-bang structure; however, it could not be solved in closed form.
A fully numerical approach was therefore suggested in \cite{smoothbangbang} by using a smoothing function to approximate the optimal bang-bang thrust profile.
One limitation of these solutions is their substantial numerical components. In addition, due to their bang-bang structure, they are more difficult to implement and susceptible to uncertainties and errors because the controller is saturated at the end of the maneuver, leaving no margin for corrections, which may potentially lead to a collision or a miss. These limitations led to the simplification approach in \cite{Rethinking}, which proposed constant and linear thrust profiles, thereby avoiding the bang-bang structure at the expense of slightly reduced performance. In \cite{Lu2026}, a continuous thrust profile is obtained using a regularization technique with a guaranteed propellant efficiency compromise, but the proposed solution still has substantial numerical components and does not guarantee ground collision avoidance. 
Another approach to solving the bounded problem while accommodating additional path constraints is convex optimization. In \cite{LosslessConvex}, the authors present a bounded-thrust minimum-propellant formulation that incorporates constraints such as maximum speed limits, glideslope cone bounds, and thrust-vector orientation limitations. Because the lower thrust bound is non-convex, the problem is reformulated under additional assumptions to obtain a convex relaxation. The resulting approach provides a computationally efficient numerical solution. However, unlike the present work, it does not yield an analytical-based guidance law and relies on the online solution of an optimization problem. 

None of the analytically based guidance laws presented thus far explicitly addressed ground collision. This limitation was addressed in \cite{wong2006guidance}, which divided the problem into two parts and introduced an intermediate point,  
just above the landing site, using the solution to the APDG two-point boundary problem \cite{APDG}. However, this solution is not optimal.
Both \cite{OptimalFeedbackIP} and \cite{ZEMZEV} also introduced an intermediate point to prevent ground collision, and the acceleration profiles in those solutions were piecewise-optimal solutions to the LQ problem. However, the proposed solutions still divided the problem into two separate phases (before and after the intermediate point) and had substantial numerical components. In \cite{OrandVitaly}, an analytical optimal-control-based LQ powered descent guidance law was proposed, 
which enables a soft landing while passing through an optimally selected intermediate point.
The optimal intermediate point was selected such that it minimized a single cost function, shaped the trajectory to avoid a ground collision, and controlled the lander’s terminal approach direction. However, this guidance law does not account for controller bounds. Interestingly, E-guidance, which is the optimal minimal control effort LQ solution in a constant gravitational field, guarantees that the lander does not collide with the ground if the scenario duration remains below a threshold determined by the initial conditions \cite{ZEMZEV}. However, this result also does not account for the controller bounds. If the controller is partially saturated at the threshold final time, the lander may still collide with the ground. 
Reducing the final time below the threshold only increases the required accelerations,
whereas extending the time reduces the saturation time or prevents it altogether, but causes ground collision.

Summarizing these facts, it is evident that the stringent weight limitations in space applications necessitate efficient fuel consumption.
Therefore, optimal-control-based solutions to the powered descent guidance problem are essential. Thrust limitations and ground-collision avoidance should be accounted for to guarantee a feasible solution. Moreover, analytical-based continuous solutions are particularly desirable, as they are simpler to implement on the lander's hardware.

Motivated by these observations, this paper presents a new approach to the bounded-thrust powered-descent problem with ground-collision avoidance. The proposed guidance law is derived from a linear-quadratic optimal-control formulation that incorporates a time-varying thrust-acceleration bound and a quadratic cost on the terminal miss, terminal velocity error, and thrust-acceleration control effort. The mass consumption is approximated as a time-dependent function and used to express the thrust bound as a thrust-acceleration bound.
The key idea is a hierarchical separation of the thrust allocation. The mass approximation is used to constrain the horizontal thrust-acceleration profile while keeping the vertical thrust-acceleration unbounded, thereby eliminating the coupling between the horizontal and vertical channels. This separation enables analytical ground-collision avoidance under bounded thrust and yields explicit collision-avoidance conditions. Unlike existing bounded-thrust powered-descent approaches that typically rely on numerical optimization and trajectory-shaping constraints, such as glide-slope or thrust-orientation limits, the proposed method derives these conditions analytically.
Building on this formulation, the guidance law predicts the switching times between saturated and unsaturated arcs and shapes the thrust-acceleration profile to achieve a soft landing, even when the controller remains saturated over extended portions of the trajectory. Owing to its analytical nature, the guidance law is computationally efficient, and its continuous thrust-acceleration profile facilitates real-time implementation. An earlier version of the proposed guidance law was presented by the authors in \cite{nataf2026linear}.

The remainder of this paper is organized as follows. The mathematical models are introduced in \secref{sec:ModelsDerivation}. The optimization problem formulation and the coordinate transformation used in the derivation are described in \secref{sec:OptimizationProblemFormulationAndOrderReduction}, followed by the guidance law derivation in \secref{sec:OWBPDG_GuidanceLawDerivation}. The implementation of the proposed algorithm is then discussed in \secref{APP:IterativeScheme}, followed by the performance analysis in \secref{sec:PerformanceAnalysis}. Finally, the conclusions are given in \secref{sec:Conclusions}, and technical derivations are provided in the Appendices.

\section{Models Derivation} \label{sec:ModelsDerivation}
The kinematics, in a non-rotating inertial Cartesian system $(x,y,z)$ attached to the required terminal position on the flat surface of the planet, is
\begin{subequations} \label{eq:motion_eqn}
\begin{align} \label{r_dot}
  \dot{\bf r}  & =  {\bf V}  \\ \label{v_dot}
  \dot{\bf V}  & = {\bf u} + {\bf g}, \quad   {\bf u}  = {\bf T}/m 
 \\ \label{u_from_T} 
      \dot{m} &= -\frac{\Vert\bf{T} \Vert}{I_{sp}g_0} 
\end{align}
\end{subequations}
where ${{\bf r}= \left[r_x,r_y,r_z \right]^T}$ is the position vector of the lander, ${{\bf V}= \left[V_x,V_y,V_z \right]^T}$ is the velocity vector of the lander, $ {\bf g}  = \left[0,0,g_z \right]^T $ is the gravitational acceleration, ${\bf u}  = \left[u_x, u_y, u_z \right]^T$ is the thrust-acceleration control command, ${\bf T} = \left[T_x,T_y,T_z \right]^T $ denotes the thrust, $I_{sp}$ is the specific impulse, $g_0$ is the standard gravitation acceleration on the Earth, and $m$ is the lander's mass. 

\begin{figure} [htbp]
 \centering
    \includegraphics[scale  = 0.65]{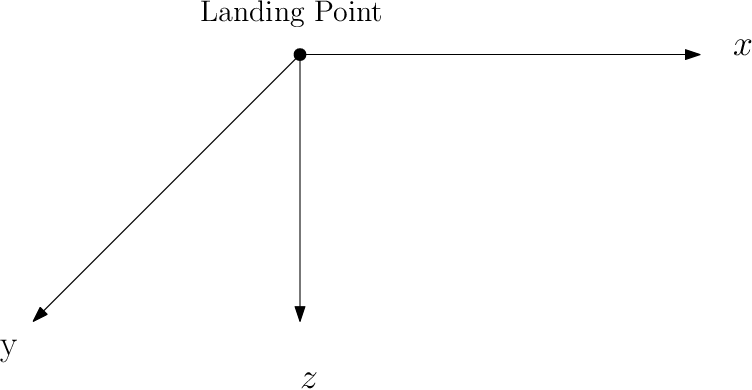}
        \caption{Problem geometry.}
    \label{fig:geometry}
\end{figure}

The origin of the coordinate system is defined in the required landing point, with the $z$-axis pointing downwards. The $x$-axis is in the down-range direction, and the $y$-axis is in the cross-range direction (see Fig. \ref{fig:geometry}).

The Equations of Motion (EOM) of the lander in \eqref{eq:motion_eqn} depend on the mass and the thrust. Since the mass rate depends on the required thrust (\eqref{u_from_T}), the mass dynamics are coupled with the thrust-acceleration.
In order to eliminate the coupling in the EOM, and to derive a simpler analytical solution for the controller, the mass will be assumed to be a known positive function $m(t_{go})>0$, where $t_{go}=t_f-t$ is the time-to-go, $t$ is time, and $t_f$ is the final time.
The guidance law implementation suggests an iterative process to approximate the mass as a function of time. The detailed description of the mass approximation is presented in subsection \ref{5.1_MassApproximation}. 

Under this assumption, the state vector of the problem is 
\begin{equation} \label{state_vec}
{\bf x}  = \begin{bmatrix}
            {\bf r}^T & {\bf V}^T 
        \end{bmatrix}^T
\end{equation}
Using the EOM in \eqref{eq:motion_eqn}
we can, therefore, represent the lander kinematics by
\begin{equation} \label{dyn_ss}
            {\dot{\bf{x}}} = {\bf Ax+Bu+Cg} 
\end{equation}
where
\begin{equation} \label{mat_ABC}
{\bf A} = \begin{bmatrix}
{[{\boldsymbol 0}]}_{3 \times 3} & {[{\boldsymbol I}]}_{3 \times 3} \\
{[{\boldsymbol 0}]}_{3 \times 3} & {[{\boldsymbol 0}]}_{3 \times 3}
\end{bmatrix}, \quad
{\bf B} = \begin{bmatrix}
{[{\boldsymbol 0}]}_{3 \times 3} \\
{[{\boldsymbol I}]}_{3 \times 3} 
\end{bmatrix}, \quad
{\bf C} = \begin{bmatrix}
{[{\boldsymbol 0}]}_{3 \times 3} \\
{[{\boldsymbol I}]}_{3 \times 3} 
\end{bmatrix} 
\end{equation}
and $[{\boldsymbol 0}]$ and $[{\boldsymbol I}]$ denote a matrix of zeros and the identity matrix, respectively, in the specified dimensions.

\section{Optimization Problem Formulation and Coordinate Change} \label{sec:OptimizationProblemFormulationAndOrderReduction}

\subsection{Optimization Problem Formulation} \label{subsec:OptimizationProblemFormulation}
The cost function we would like to minimize is given by
\begin{equation} \label{cost_func}
\mathcal{J} = \frac{1}{2} \Vert{{\bf M}{\bf x} (t_f)}\Vert ^2 + 
\frac{1}{2} \int_{0}^{t_f} {{\Vert {\bf u}\Vert }^2} \ dt, \quad s.t.  \quad 
\Vert{\bf T}\Vert \leq T_{\max}
\end{equation}
where $T_{\max}$ is the maximum thrust, and ${\bf M}$ is a diagonal weighting matrix in which ${\bf M}_{rr}$ weights the terminal position errors and ${\bf M}_{VV}$ weights the terminal velocity errors.
\begin{equation} \label{M_def}
{\bf M} = \left[{\bf M}\right]_{6\times 6} = diag\{\left[{\bf M}_{rr}\right]_{3\times 3},\left[{\bf M}_{VV}\right]_{3\times 3} \}
\end{equation}
The cost function weights the thrust-acceleration control effort and the deviation of the state vector from the desired terminal conditions, which in the present case are $[{\bf 0}]_{6\times1}$. The terminal conditions are imposed softly through non-negative weights that penalize the terminal position and velocity errors. In the limit as the weights tend to infinity, the soft terminal constraints become hard terminal constraints.

Assuming the mass  as a function of time is known and defining a positive function $f(t_{go})$
\begin{equation}\label{eqn:F}
    f(t_{go}) = \frac{1}{m(t_{go})}> 0
\end{equation}
The problem can be reformulated using \eqref{v_dot}
\begin{equation} 
    \mathcal{J} = \frac{1}{2} \Vert{{\bf M}{\bf x} (t_f)}\Vert ^2 +  \frac{1}{2} \int_{0}^{t_f} {{\Vert {\bf u}\Vert }^2} \ dt, \quad s.t. \quad 
    \Vert{\bf u}\Vert \leq f(t_{go}) \cdot {T_{\max}}
\end{equation}
Assuming $|{u}_z| \leq {T_{\max}}\cdot f(t_{go}) \ \ \forall t$, and has priority over ${u}_x$ and ${u}_y$, the formulation of the problem is 
\begin{equation} \label{eqn:costfunctionwithu}
    \mathcal{J} = \frac{1}{2} \Vert{{\bf M} {\bf x} (t_f)}\Vert ^2 +  \frac{1}{2} \int_{0}^{t_f} {{\Vert {\bf u}\Vert }^2} \ dt, \quad s.t. \quad 
    \Vert{\bf u}_{xy} \Vert \leq h_{\max}(t_{go}) \cdot {T_{\max}}, \quad {\bf u}_{xy}=\begin{bmatrix}
    u_{x} & 
    u_{y}  
    \end{bmatrix}^T
\end{equation}
where $h_{\max} (t_{go})$ is assumed to be a known bounding function given by
\begin{equation} \label{eqn:gdef}
    h_{\max} (t_{go}) =  \sqrt{f^2(t_{go}) - \frac{u^2_z(t_{go})} {T^2_{\max}}}
\end{equation}

The motivation for this problem formulation might not be clear at first glance. It stems from the fact that ground collision in the unsaturated case only depends on the motion in the $z$-axis, as presented in \cite{ZEMZEV}. In such a scenario, if the flight time $t_f$ is smaller than some maximal value $t_f<t_{f_{\max}}$, which depends on the initial conditions in the $z$-axis, then ground collision is avoided. Thus, assuming  $|{u}_z| \leq {T_{\max}}\cdot f(t_{go})$ and prioritizing the $z$-axis will prevent ground collision, assuming the trajectory in the $x$-$y$ plane can still be brought to the final conditions in the allotted time $t_f<t_{f_{\max}}$ with the remainimg thrust. 
%
\subsection{Coordinate Change}  \label{subsec:OrderReduction}
We use a transformation sometimes called \emph{terminal projection} \cite{bryson2018applied}. This transformation significantly simplifies the optimal control solution. Let
\begin{equation} \label{Z_def}
    {\bf Z}(t) \triangleq {\begin{bmatrix}
    [{{\bf Z}}_r^T(t)]_{1 \times 3} & 
    [{{\bf Z}}_V^T(t)]_{1 \times 3}  
    \end{bmatrix}}^T=
    {\bf M}{\boldsymbol \phi}(t_f,t){\bf x}(t) + {\bf M}\int_{t}^{t_f} {\boldsymbol \phi}(t_f,\tau) \  {\bf C}{{\bf g}} \ d\tau
\end{equation}
where ${\bf \Phi}(t_f,t)$ is the transition matrix associated with \eqref{dyn_ss} from $t$ to $t_f$. The vector ${\bf Z}(t)$, referred to as the Weighted Zero-Effort Vector, has a clear physical interpretation. The component ${\bf Z}_r(t)$ is the Weighted-Zero-Effort-Miss (WZEM), while ${\bf Z}_V(t)$ is the Weighted-Zero-Effort-Velocity (WZEV). They represent the weighted terminal position and velocity errors, respectively, that would result if no further control were applied, where the weighting is defined by the matrix ${\bf M}$.

Differentiating \eqref{Z_def} with respect to time, substituting \eqref{dyn_ss}, and the following transition matrix property
\begin{equation} \label{trans_prop}
\dot{\bfphi}(t_f,t) = -\bfphi(t_f,t)\cdot\boldsymbol{A} \  ,  \quad \quad \boldsymbol{\phi}(t,t) = {[{\boldsymbol I}]}_{6 \times 6} 
\end{equation}
We get
\begin{equation} 
 \label{z_dot_with_b_tilde}
 \dot{\bf Z}(t)=  \Tilde{\bf B}(t_{go}) {\bf u}(t), \quad
 {\Tilde{\bf B}}(t_{go}) = {\bf M} {\boldsymbol \phi} (t_f,t) {\bf B}=
 \begin{bmatrix}
    {\bf M}_{rr} t_{go} \\
    {\bf M}_{VV}
 \end{bmatrix}
\end{equation}
Where the transition matrix ${\boldsymbol \phi}(t_f,t)$ is derived in \cite{Gutman} 
and equals 
\begin{equation} \label{eqn:mrx_Tras}
    {\boldsymbol \phi}(t_f,t) =
{\begin{bmatrix}
     {[{\boldsymbol I}]}_{3 \times 3} & t_{{go}}\cdot {[{\boldsymbol I}]}_{3 \times 3} \\ 
    {[{\boldsymbol 0}]}_{3 \times 3}  & {[{\boldsymbol I}]}_{3 \times 3} 
    \end{bmatrix}}
\end{equation}
Using the fact that according to \eqref{Z_def} ${\bf Z}(t_f)={\bf M}{\bf x}(t_f)$ the cost function in \eqref{eqn:costfunctionwithu} can now be expressed as
\begin{equation} \label{eqn:costfunction1}
 \mathcal{J} =\frac {1}{2}{\Vert {\bf Z}(t_f) \Vert}^2 + \frac{1}{2} \int_{0}^{t_f} {{\Vert {\bf u}\Vert }^2} \ d{t}, \quad s.t. \quad 
 \Vert{\bf u}_{xy} \Vert  \leq h_{\max}(t_{go}) \cdot {T_{\max}}
\end{equation}
For brevity, we omit the time dependence of the variables in the remainder of the paper unless it is necessary to understand the derivation.
%
\section{Bounded Guidance Law Derivation} \label{sec:OWBPDG_GuidanceLawDerivation}
\subsection{General Solution} \label{subsec:OWBPDG_GeneralSolution}
The augmented Hamiltonian of the bounded problem is
\begin{equation} \label{newhamiltonian}
{\mathcal{H}_B} = \mathcal{H}+ \mu \left[ \Vert{{\bf u}}_{xy}\Vert ^2 -  {h_{\max}^2(t_{go})} \cdot {T^2_{\max}} \right], \quad \mathcal{H}= \frac{1}{2}{\Vert {\bf u} \Vert ^2}  + {\boldsymbol \lambda}^T { \Tilde{\bf B}} {\bf u}, \quad {\boldsymbol \lambda}^T 
    = \begin{bmatrix}
    {\boldsymbol{{\lambda}}}_r^T & {\boldsymbol{{\lambda}}}_V^T
    \end{bmatrix}
\end{equation}
where $\mathcal{H}$ is the unbounded Hamiltonian, $\mu$ is a Lagrange multiplier, and $\boldsymbol{\lambda}$ are the co-states.

 The adjoint equations and their boundary conditions at $t=t_f$ yield  \begin{equation} 
 \label{adj_eq}
     {\dot{\boldsymbol \lambda}}^T =  -\frac{\partial {\mathcal{H}}_B}{\partial {\bf Z}} = -\frac{\partial \mathcal{H}}{\partial {\bf Z}}= {\bf 0}^T, \quad  
     {\boldsymbol{\lambda}}^T(t_f) = {{\bf Z}^T(t_f)}  \quad  \Rightarrow \quad
      {\boldsymbol \lambda} (t) = {\boldsymbol \lambda} (t_f) = {{\bf Z}(t_f)}
 \end{equation}
 The optimal unbounded controller ${\bf u}^*$ satisfies \cite{Gutman}
 \begin{equation} \label{eq:unbounded_optimal_controller}
     \frac{\partial \mathcal{H}}{\partial {\bf u}} = {\bf 0}^T \quad \Rightarrow \quad
     {\bf u}^* = - \tilde{\bf B}^T(t_{go}) {\bf Z}(t_f)=
 -{\bf M}_{rr}{\bf Z}_{r}(t_f)\cdot t_{go} -{\bf M}_{VV}{\bf Z}_{V}(t_f)
 \end{equation}
While the optimal controller for the bounded problem when the control constraint is inactive has the same structure
\begin{equation} \label{eq:optimal_controller_general_bounded_unsaturated}
     \frac{\partial \mathcal{H}}{\partial {\bf u}} = {\bf 0}^T \quad \Rightarrow \quad \tilde{\bf u} = -\tilde{\bf B}^T(t_{go}) {\bar{\bf Z}}(t_f) = 
-{\bf M}_{rr}{\bar{\bf Z}}_{r}(t_f)\cdot t_{go} -{\bf M}_{VV} {\bar{\bf Z}}_{V}(t_f)
\end{equation} 
where $\bar{\bf Z}(t_f)$ is the terminal weighted Zero-Effort-Vector of the bounded problem, which might be different than that of the unbounded problem ${\bf Z}(t_f)$ and is, therefore, denoted with a bar. To prevent confusion, all the notations in the rest of the paper that explicitly refer to the bounded problem and might be confused with the unbounded problem are denoted with a bar.

The optimal controller when the control constraint is active is obtained when all partial derivatives of $\mathcal{H}_{B}$ are zero (see \cite{bryson2018applied})
\begin{subequations}
    \label{eqn:HBLambdaB}
    \begin{equation} \label{eqn:HB_Normu}
        \frac{\partial \mathcal{H}_{B}}{\partial {\mu}} = {0} = \Vert{\bar {\bf u}}_{xy}\Vert ^2 -{h_{\max}^2(t_{go})} \cdot {T^2_{\max}} 
        \quad \Rightarrow \quad
        \Vert{\bar {\bf u}}_{xy} \Vert ^2 = {h_{\max}^2(t_{go})} \cdot {T^2_{\max}}
    \end{equation}
\begin{equation}
\label{eqn:HBu} 
    \frac{\partial \mathcal{H}_{B}}{\partial \bar{\bf u}} = 
    {\bf 0}^T \quad \Rightarrow \quad
    \begin{cases} 
    \displaystyle
        \frac{\partial \mathcal{H}_{B}}{\partial \bar{\bf u}_{xy}}=
       {\bf 0}^T 
        \quad &\Rightarrow \quad \displaystyle
        \bar{\bf u}_{xy}=\frac{\tilde{\bf u}_{xy}}{1+2\mu}
    \\ \displaystyle
        \frac{\partial \mathcal{H}_{B}}{\partial \bar{u}_z}=
        0 \quad &\Rightarrow \quad \bar{u}_z=\tilde{u}_z
    \end{cases} 
\end{equation}
\end{subequations}
where $\tilde{\bf u}_{xy}=\begin{bmatrix} \tilde u_{x} & \tilde u_{y} \end{bmatrix}^T$ and $\tilde u_z$ denote the $x$-$y$ plane and $z$-axis components, respectively, of the control law in \eqref{eq:optimal_controller_general_bounded_unsaturated}.
Substituting the optimal $\bar{\bf u}_{xy}$ controller in \eqref{eqn:HBu} into \eqref{eqn:HB_Normu} 
\begin{equation}
    1+2\mu 
     = \pm \frac{\Vert\tilde{\bf u}_{xy}\Vert}{h_{\max}(t_{go}) \cdot T_{\max}}
\end{equation}
From the second-order condition $(1+2\mu)>0$
\begin{equation} 
    \frac{{\partial^2 \mathcal{H}_{B}}}{\partial \bar{\bf u}_{xy}^2} = \left (1+2\mu \right ) {[{\boldsymbol I}]}_{2 \times 2} > 0 
\end{equation}
Therefore, the optimal bounded controller when the control constraint is active equals 
\begin{equation}
\label{eqn:BoundedOptimalController}
    \bar{\bf u}_{xy} = \frac{\tilde{\bf u}_{xy}}{\Vert \tilde{\bf u}_{xy} \Vert}
   \ h_{\max}(t_{go})\cdot T_{\max}
\end{equation}
The optimal controller for the bounded problem is, therefore
\begin{subequations} \label{eqn:controller}
\begin{equation}
    \bar{u}^*_z = \tilde{u}_z
\end{equation}
\begin{equation} 
    \bar{\bf u}_{xy}^* = \begin{cases}
    \tilde{\bf u}_{xy} 
            &
            \Vert \tilde{\bf u}\Vert \leq f(t_{go}) \cdot T_{\max} \\
             \frac{\tilde{\bf u}_{xy}}{\Vert{\tilde{\bf u}_{xy}}\Vert} \ h_{\max}(t_{go})\cdot T_{\max}
             &
             \Vert \tilde{\bf u}\Vert > f(t_{go}) \cdot T_{\max} 
    \end{cases}
\end{equation}
\end{subequations}
where $\tilde{\bf u}$ is the unsaturated optimal controller in the bounded problem in \eqref{eq:optimal_controller_general_bounded_unsaturated}. Note that the optimal controller in the $z$-axis is decoupled from the optimal controller in the $x$-$y$ plane.
\subsection{Switching Times}
To obtain the optimal controller, it is necessary to identify the switching times between intervals with active control constraints and those with inactive constraints.
The switching times are defined using the intersection points between the unsaturated controller and its time-dependent bound. For simplicity, the problem is analyzed in the time-to-go domain.

Let $S_{\max}(t_{go})$ be the following time-varying function
\begin{equation}
     S_{\max}(t_{go}) = \Vert \tilde{\bf u} \Vert ^2- f^2(t_{go}){T^2_{\max}}  
\end{equation}
such that the switching time are the roots of $S_{\max}(t_{go})$.
For simplicity, in the rest of the derivation we assume ${\bf M }= \alpha{[{\boldsymbol I}]}_{6 \times 6}$. 
Substituting \eqref{eq:optimal_controller_general_bounded_unsaturated} 
into $S_{\max}(t_{go})$ yields
 \begin{equation}
    \label{eqn:Stgo_generic}
      S_{\max}(t_{go}) =  \alpha^2 \left[ \Vert{\bar{\bf Z}_{r}}(t_f) \Vert^2\cdot t^2_{go} + 2\bar{\bf Z}^T_{{{r}}}(t_f)\bar{\bf Z}_{{V}}(t_f)\cdot t_{go} + \Vert{\bar{\bf Z}_{{V}}(t_f)}\Vert^2 \right]- f^2(t_{go}){T^2_{\max}}
 \end{equation} 
 Differentiating with respect to $t_{go}$
\begin{align}
    S'_{\max}(t_{go}) &= 2\alpha^2 \left[ \Vert{\bar{\bf Z}_{{r}}(t_f)}\Vert^2\cdot t_{go} + \bar{\bf Z}^T_{{{r}}}(t_f)\bar{\bf Z}_{{V}}(t_f) \right]-2 f(t_{go}) f'(t_{go})T^2_{\max} 
\end{align}

Since we solve the mathematical problem of finding roots (i.e., $S_{\max}(t_{go}) = 0$), non-physical solutions may be obtained.
These solutions correspond to cases in which the controller intersects its bounds either before the start or after the end of the scenario.

\begin{figure}
    \centering
    \includegraphics[width=0.4\linewidth]{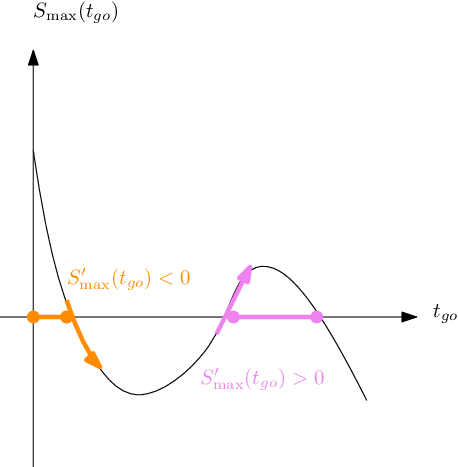}
    \caption{Conditions for determining the constrained intervals.}
    \label{fig:Conditions for determining the bounded intervals}
\end{figure}

Let
$\mathbb{R}_{B} $ be the time intervals in which the 
constraints of the controller are active (i.e., the controller is saturated)
    \begin{equation}\label{eqns:IntersectionsInterval}
      \mathbb{R}_{B} = \left\{ t_{go}  \in \mathbb{R}^1 \ | \  S_{\max}(t_{go}) > 0 \ ,\ \   0 \leq t_{go} \leq t_f \right\}
    \end{equation}
Moreover, let us define the ordered set of intersection times with the 
thrust bound as
    \begin{equation} \label{eqns:Intersections}
       t^{B}_{go} = \left\{t_{go}  \in \mathbb{R}^1 \ | \  S_{\max}(t_{go}) = 0 \ ,\ \   0 \leq t_{go} \leq t_f \right\}
    \end{equation}
Let
$t^{B}_{{go}_i}$ be elements in the set $t^{B}_{go}$ 
such that
\begin{equation}
        t^{B}_{{go}}  = \bigcup_i  t^{B}_{{go}_i},  \quad t^{B}_{{go}_i}\le t^{B}_{{go}_{i+1}} \quad \forall \ i \in \left\{ 1,2,...,\lvert t^{B}_{go} \rvert \right\}
\end{equation}
with the corresponding time intervals in which the control constraint is active 
and $\mathbb{R}_{{B}_i}$
such that $        \mathbb{R}_{B} = \bigcup _{i} \mathbb{R}_{{B}_i} 
$.
The derivative of $S_{\max}(t_{go})$ in the switching times defines the time intervals in which the control constraint is active. 
For an 
intersection at $t^{B}_{{go}_i}$, the time interval $\mathbb{R}_{{B}_i} $ is 
\begin{subequations} \label{eqns:conditions_for_time_intervals}
  \begin{equation}
     \mathbb{R}_{{B}_i} = \left[t^{B}_{{go}_{i-1}},t^{B}_{{go}_{i}} \right] \iff S'_{\max}(t^{B}_{{go}_i}) < 0 
\end{equation}  
\begin{equation}
     \mathbb{R}_{{B}_i} = \left[t^{B}_{{go}_{i}},t^{B}_{{go}_{i+1}} \right] \iff S'_{\max}(t^{B}_{{go}_i}) \geq 0 
\end{equation}
\end{subequations} 
Time segments that contain time intervals of $t_{go}<0$ or $t_{go}>t_f$ should be truncated at $0$ or $t_f$, respectively.
Although $\mathbb{R}_{{B}_{i}}$ and $\mathbb{R}_{{B}_{i+1}}$ can be identical, in the union, there are no repetitions. 
The interval in which the constraints are inactive is defined by $\mathbb{R}_{NB}=[0,t_f] \backslash \mathbb{R}_{B}$. 
Figure \ref{fig:Conditions for determining the bounded intervals} presents a graphical description of the conditions for determining the constrained intervals based on the switching times and $S'_{\max}(t_{go})$ (i.e., the slope of $S_{\max}(t_{go})$).  
\subsection{Bounded Open-loop Controller} \label{subsubsec:openloopcontrollerforgeneric}
The optimal controller obtained in \eqref{eqn:controller} is expressed using the intervals ${\mathbb{R}_{B}}$ and ${\mathbb{R}_{NB}}$ as
\begin{subequations} \label{eqn:optimalcontrollercases}
\begin{equation}\label{eqn:optimalcontrollercases_z}
    \bar{u}_z^* = \tilde{u}_z \qquad t_{go} \in  [0,t_f] 
\end{equation}
\begin{equation} \label{eqn:optimalcontrollercases_zy}
    \bar{\bf u}_{xy}^* = \begin{cases}
      \tilde{\bf u}_{xy} 
            &
            t_{go} \in  {\mathbb{R}_{NB}}\\
             \frac{\tilde{\bf u}_{xy}}{\Vert{\tilde{\bf u}_{xy}}\Vert} \ h_{\max}(t_{go})\cdot T_{\max}
             &
             t_{go} \in  {\mathbb{R}_{B}} 
    \end{cases}
\end{equation} 
\end{subequations}
The optimal controller is a function of $\bar{\bf Z}(t_f)$, therefore, we substitute the \eqref{eqn:optimalcontrollercases} into \eqref{z_dot_with_b_tilde} and integrate from $0$ to $t_f$
\begin{equation}
    \bar{\bf Z}(t_f)- {\bar{\bf Z}}(0) =  \int_{0}^{t_f} \dot{\bar{\mathbf{Z}}}(t) \,dt     = 
     \int^{t_{f}}_{0} \dot{\bar{\mathbf Z}}(t_{go}) \,dt_{go}
\end{equation}
Since the $z$-axis is fully decoupled from the other axes. Let us define the components of $\bar{\bf Z}$
\begin{equation}
     \bar{\bf Z}_r= \begin{bmatrix}
       \bar{\bf Z}_{r_x} & \bar{\bf Z}_{r_y} & \bar{\bf Z}_{r_z}
   \end{bmatrix}^T, \quad 
     \bar{\bf Z}_V= \begin{bmatrix}
       \bar{\bf Z}_{V_x} & \bar{\bf Z}_{V_y} & \bar{\bf Z}_{V_z}
   \end{bmatrix}^T
\end{equation}
where the $x$-$y$ plane and $z$-axis components of $\bar{\bf Z}$ are
\begin{equation}
     \bar{\bf Z}_{xy}= \begin{bmatrix}
        \bar{\bf Z}_{r_{xy}}^T & \bar{\bf Z}_{V_{xy}}^T
   \end{bmatrix}^T, \quad
   \bar{\bf Z}_{r_{xy}}= \begin{bmatrix}
       \bar{\bf Z}_{r_x} & \bar{\bf Z}_{r_y}
   \end{bmatrix}^T, \quad
   \bar{\bf Z}_{V_{xy}}= \begin{bmatrix}
       \bar{\bf Z}_{V_x} & \bar{\bf Z}_{V_y}
   \end{bmatrix}^T, \quad
     \bar{\bf Z}_z= \begin{bmatrix}
       \bar{\bf Z}_{r_z} & \bar{\bf Z}_{V_z}
   \end{bmatrix}^T 
\end{equation}
and the dynamics of $\bar{\bf Z}$ in \eqref{z_dot_with_b_tilde} can be partitioned into independent $x$-$y$ plane and $z$-axis subsystems
\begin{equation} 
 \label{z_dot_with_b_tilde_zy_z}
 \dot{\bar{\mathbf Z}}_{xy}=  \Tilde{\bf B}_{xy} \bar{\bf u}_{xy}, \quad
 {\Tilde{\bf B}}_{xy}(t_{go}) = \alpha
 \begin{bmatrix}
     {[{\boldsymbol I}]}_{2 \times 2} \cdot t_{go} \\
      {[{\boldsymbol I}]}_{2 \times 2}
 \end{bmatrix}, \quad 
 \dot{\bar{\mathbf Z}}_{z}=  \Tilde{\bf B}_{z} \bar{\bf u}_{z}, \quad
 {\Tilde{\bf B}}_{z}(t_{go}) = \alpha
 \begin{bmatrix}
     t_{go} \\
      1
 \end{bmatrix}
\end{equation}
Substituting \eqref{eqn:optimalcontrollercases_z}, integrating along the $z$-axis, and rearranging 
yields
 \begin{equation}\label{eqn:IntegralGeneralExp_Z}
     \bar{\bf Z}_{z}(t_f)={\bf P}_z \bar{\bf Z}_{z}(0), \quad {\bf P}_z=\boldsymbol{\Delta}_z^{-1}\begin{bmatrix}
    1 + \alpha^2 t_{go} 
     &
     -\frac{\alpha^2}{2}t^2_{go} 
     \\ 
    - \frac{\alpha^2}{2}t^2_{go} 
     &
      1 + \frac{\alpha^2}{3}t^3_{go} 
 \end{bmatrix}, \quad \boldsymbol{\Delta}_z=\left(1 + \frac{\alpha^2}{3}t^3_{go}  \right) \left(1 + \alpha^2 t_{go} \right)- \frac{\alpha^4}{4}t^4_{go}  
\end{equation}
Where $\boldsymbol{\Delta}_z>0 \; \forall t_{go}\ge0$.
\begin{rem} 
Assuming that $|\bar{u}_z| \leq {T_{\max}}\cdot f(t_{go})$, and since the $z$-axis controller is decoupled from the controller in the $x$-$y$ plane, it follows that $\bar{\bf Z}_{z}(t_f)={\bf Z}_{z}(t_f)$. Hence, the $z$-axis trajectory is identical in the bounded and unbounded cases.
It was shown in \cite{ZEMZEV} that ground collision is avoided if and only if the terminal position corresponds to a maximum point of the $z$-axis trajectory, or equivalently, if the terminal acceleration in the $z$ direction is negative. This yields the following condition
\begin{equation} \label{eq:t_max}
     {t_f} < {t_f}_{\max}=\frac{-3{r}_z(0)}{V_z(0)}
\end{equation}
Note that this condition was originally derived for the unbounded controller under the assumption that $V_z \cdot r_z < 0$. When $V_z \cdot r_z \geq 0$, ground-collision avoidance is guaranteed for any $t_f$. Because the $z$-axis controller is unaffected by the thrust bound, the derivation remains valid, and the same condition applies in the bounded case.
\end{rem}

Unlike the $z$-axis, the integrand in the $x$-$y$ plane varies as a function of the switching times (i.e., active and inactive constraints); the integral is, therefore, decomposed to 
\begin{equation} \label{eqn:IntegralGeneralExp}
    \bar{\bf Z}_{{xy}}(t_f)- \bar{\bf Z}_{xy}(0) 
    = 
     \int_{\mathbb{R}_{B}} \dot{\bar{\mathbf Z}}_{xy}(t_{go}) \,dt_{go}
     +
     \int_{\mathbb{R}_{NB}}
     \dot{\bar{\mathbf Z}}_{xy}(t_{go}) \,dt_{go}
\end{equation}
Substituting \eqref{eqn:optimalcontrollercases_zy} into \eqref{z_dot_with_b_tilde_zy_z} and integrating yields the indefinite integral $\bar{\bf I}^{xy}_{NB}$, which is defined over the time intervals during which the control constraint is inactive
\begin{equation} \label{eqn:unboundedintegral}
 \bar {\bf I}^{xy}_{NB}  = -\int{\Tilde{\bf B}_{xy}(t_{go})\Tilde{\bf B}_{xy}^T(t_{go}){\bar{\bf Z}_{xy}(t_f)} }\  \,dt_{go}=
- \alpha^2  
    \begin{bmatrix}
           \frac{1}{3} t^3_{go} {[{\boldsymbol I}]}_{2 \times 2}   &
            \frac{1}{2} t^2_{go}  {[{\boldsymbol I}]}_{2 \times 2}  \\ 
            \frac{1}{2} t_{go} {[{\boldsymbol I}]}_{2 \times 2}    & 
             t_{go}{[{\boldsymbol I}]}_{2 \times 2}
        \end{bmatrix}  
        \bar{\bf Z}_{{xy}}(t_f) 
\end{equation}
In contrast, during intervals in which the control constraint is active, the indefinite integral $\bar{\bf I}^{xy}_{B}$ is used
\begin{equation} \label{eqn:boundedintegral_u}
   \bar{\bf I}^{xy}_{B} = 
    T_{\max} \int  \Tilde{\bf B}_{xy}(t_{go})
   \frac{\tilde{\bf u}_{xy}(t_{go})}{\Vert{\tilde{\bf u}}_{xy}(t_{go})\Vert}
   \ h_{\max}(t_{go}) \,dt_{go} 
\end{equation} 
Substituting $\tilde{\bf u}_{xy}$ yields
\begin{equation} 
\label{eqn:boundedintegral}
    \bar{\bf I}^{xy}_{B} =  \int    \frac{-\alpha  T_{\max} h_{\max}(t_{go})}{\sqrt{\Vert\bar{\bf Z}_{r_{xy}}(t_f)\Vert^2 t^2_{go}+2\bar{\bf Z}_{r_{xy}}^T(t_f) \bar{\bf Z}_{V_{xy}}(t_f) t_{go} + \Vert \bar{\bf Z}_{V_{xy}}(t_f) \Vert^2}} 
    \begin{bmatrix}
        t^2_{go}\bar{\bf Z}_{r_{xy}}(t_f)+ t_{go}  \bar{\bf Z}_{V_{xy}}(t_f) \\ 
        t_{go} \bar{\bf Z}_{r_{xy}}(t_f) + \bar{\bf Z}_{V_{xy}}(t_f)
    \end{bmatrix} \,dt_{go} 
\end{equation}
and \eqref{eqn:IntegralGeneralExp}, can be rewritten as
\begin{equation} \label{eqn:Ztf_Z0}
    \bar{\bf Z}_{{xy}}(t_f)- \bar{\bf Z}_{xy}(0) 
    = 
     \bar{\bf I}^{xy}_{NB}({\mathbb{R}_{NB}})+
     \bar{\bf I}^{xy}_{B}({\mathbb{R}_{B}})
\end{equation}
where $\bar{\bf I}^{xy}_{NB}({\mathbb{R}_{NB}})$ and $\bar{\bf I}^{xy}_{NB}({\mathbb{R}_{B}})$ denote the corresponding integrals evaluated over the intervals ${\mathbb{R}_{NB}}$ and ${\mathbb{R}_{B}}$, respectively.  

The resulting expressions are then substituted into \eqref{eqn:Ztf_Z0}. The switching times are ${\bar{\bf Z}_{xy}}(t_f)$ dependent; therefore, the following nonlinear system of 5 equations is obtained
\begin{subequations}
\label{eqns:NonInvertabelZtfToZt}
    \begin{equation}
    \bar{\bf Z}_{{xy}}(t_f)- \bar{\bf Z}_{xy}(0) = 
    \bar{\bf I}^{xy}_{NB}({\mathbb{R}_{NB}})+
     \bar{\bf I}^{xy}_{B}({\mathbb{R}_{B}}) 
    \end{equation}
  \begin{equation} 
      f^2({t^{B}_{{go}_i} }) {T^2_{\max}}  = \alpha^2  \left[ \Vert{\bar{\bf Z}_{r}(t_f)}\Vert^2\cdot \left(t^{B}_{{go}_i} \right)^2 + 2\bar{\bf Z}^T_{r}(t_f) \bar{\bf Z}_{V}(t_f)\cdot t^{B}_{{go}_i}  + \Vert{\bar{\bf Z}_{V}(t_f)}\Vert^2 \right]  \qquad \forall t^{B}_{{go}_i} 
      ,\quad  i \in \left[ 1,\lvert t^{B}_{go} \rvert \right]
 \end{equation} 
\end{subequations}
where
\begin{subequations}
\begin{align}
 \Vert{\bar{\bf Z}_{r}(t_f)}\Vert^2 =\Vert{\bar{\bf Z}_{r_{xy}}(t_f)}\Vert^2&+{\bar{\bf Z}_{r_z}^2(t_f)}, \quad 
 \Vert{\bar{\bf Z}_{V}(t_f)}\Vert^2=
 \Vert{\bar{\bf Z}_{V_{xy}}(t_f)}\Vert^2+\bar{\bf Z}_{V_z}^2(t_f) \\
 \bar{\bf Z}^T_{r}(t_f) \bar{\bf Z}_{V}(t_f)&=\bar{\bf Z}^T_{r_{xy}}(t_f) \bar{\bf Z}_{V_{xy}}(t_f)+\bar{\bf Z}_{r_{z}}(t_f) \bar{\bf Z}_{V_{z}}(t_f)
\end{align}
\end{subequations}
and $\bar{\bf Z}_{r_{z}}(t_f), \bar{\bf Z}_{V_{z}}(t_f)$ can be calculated in closed from from \eqref{eqn:IntegralGeneralExp_Z}.
Substituting the zeros of this system of equations into \eqref{eqn:optimalcontrollercases}, we can obtain the open-loop controller.
This solution can be applied at each time step by replacing $t=0$ with $t$ to obtain a series of open-loop controllers.

\subsection{Polynomial Approximation of $f(t_{go})$ and $h_{\max}(t_{go})$} 
\label{subsec:OWPDG_PolyApprox}
To obtain a practical solution and to simplify the problem, polynomial approximations of the continuous functions $f(t_{go})$ and $h_{\max}(t_{go})$ are employed in this section. 
Let us assume the following 
polynomial approximations for $f(t_{go})$ and $h_{\max}(t_{go})$
\begin{subequations}
\begin{equation} \label{eqn:fasNpoly}
    f(t_{go}) = \sum_{k=0}^{N_\omega} \omega_k\cdot t_{go}^k
\end{equation} 
\begin{equation} \label{eqn:hasaply}
    h_{\max}(t_{go}) = 
        \displaystyle \sum_{k=0}^{N_a} a^{\max}_k \cdot t_{go}^k
\end{equation}
\end{subequations}
To obtain the bounded open-loop controller, we first need to find the switching times.
Substituting \eqref{eqn:fasNpoly} into \eqref{eqn:Stgo_generic} yields
  \begin{equation}
  \label{eqn:stgof}
     S_{\max}(t_{go}) =  \alpha^2 \left[ \Vert{\bar{\bf Z}_r(t_f)\Vert^2 \cdot t^2_{go} + 2\bar{\bf Z}^T_{r}(t_f)\bar{\bf Z}_V(t_f)}\cdot t_{go} + \Vert{\bar{\bf Z}_V(t_f)}\Vert^2 \right]-\left (\sum_{k=0}^{N_\omega} \omega_k \cdot t^k_{go} \right)^2
    {T^2_{\max}} 
 \end{equation} 
 Therefore, $S_{\max}(t_{go}) = 0$ is a polynomial equation with at most $2N_\omega+2$ roots, and the corresponding intersection times are determined using \eqref{eqns:Intersections}. 
It should be noted that, typically, $N_\omega \geq 1$, in which case $S_{\max}(t_{go}) = 0$ is a polynomial equation with at most $2N_\omega$ roots.
The derivative of $S_{\max}(t_{go})$ with respect to the time-to-go is required to identify the intervals during which the control constraint is active and inactive
\begin{align} 
  S'_{\max}(t_{go}) = 2\alpha^2 \left[ \Vert{\bar{\bf Z}_r(t_f)}\Vert^2\cdot t_{go} + \bar{\bf Z}^T_r (t_f)\bar{\bf Z}_V(t_f) \right]-2 T^2_{\max} \sum_{k=0}^{N_\omega} \omega_k \cdot t^k_{go} \sum_{i=0}^{N_\omega-1} (i+1) \omega_{i+1}  \cdot t^{i}_{go}  
\end{align}
The time intervals during which the control constraint is active and inactive are then obtained from \eqref{eqns:IntersectionsInterval} and \eqref{eqns:conditions_for_time_intervals}.
Substituting \eqref{eqn:hasaply} into \eqref{eqn:optimalcontrollercases} we get the open loop bounded controller
\begin{subequations}
\label{eqn:foptimalcontrollercases}
\begin{equation}
    \bar{u}_z = \tilde{u}_z \qquad t_{go} \in  [0,t_f] 
\end{equation}
\begin{equation} 
    \bar{\bf u}_{xy}^* = \begin{cases}
            \tilde{\bf u}_{xy} 
            &
            t_{go} \in  {\mathbb{R}_{NB}}
        \\ 
             \frac{\tilde{\bf u}_{xy}}{\Vert{\tilde{\bf u}_{xy}}\Vert} \ \displaystyle \sum_{k=0}^{N_a} a^{\max}_k\cdot t_{go}^k\cdot T_{\max} 
             &
             t_{go} \in  {\mathbb{R}_{B}} 
    \end{cases}
\end{equation} 
\end{subequations}

The closed-form expression for the unconstrained integral is given in \eqref{eqn:unboundedintegral}, while that of the constrained integral is obtained by substituting \eqref{eqn:hasaply} into \eqref{eqn:boundedintegral}
\begin{equation}
\label{eqn:fboundedintegral}
  \begin{split}
    \bar{\bf I}^{xy}_{B} =  \int    \frac{\displaystyle - \alpha T_{\max}\sum_{k=0}^{N_a} a_k\cdot t_{go}^k}{\sqrt{\Vert\bar{\bf Z}_{r_{xy}}(t_f)\Vert^2 t^2_{go}+2\bar{\bf Z}_{r_{xy}}^T(t_f) \bar{\bf Z}_{V_{xy}}(t_f) t_{go} + \Vert \bar{\bf Z}_{V_{xy}}(t_f) \Vert^2}} 
    \begin{bmatrix}
        t^2_{go}\bar{\bf Z}_{r_{xy}}(t_f)+ t_{go}  \bar{\bf Z}_{V_{xy}}(t_f) \\ 
        t_{go} \bar{\bf Z}_{r_{xy}}(t_f) + \bar{\bf Z}_{V_{xy}}(t_f)
    \end{bmatrix} \,dt_{go} 
\end{split}  
\end{equation}
This integral admits an analytical solution. The closed-form expression corresponding to a second-order polynomial approximation is presented in Appendix A.

Substituting the constrained and unconstrained integral expressions into \eqref{eqns:NonInvertabelZtfToZt} reduces the equation set to
\begin{subequations} \label{eqns:ZandST}
    \begin{equation} \label{eqns:NonInvertabelZtfToZt_fpoly2nd}
    \bar{\bf Z}_{{xy}}(t_f)- \bar{\bf Z}_{xy}(0) = 
    \bar{\bf I}^{xy}_{NB}({\mathbb{R}_{NB}})+
     \bar{\bf I}^{xy}_{B}({\mathbb{R}_{B}}) 
    \end{equation}
  \begin{equation} \label{eqn:2ndorderpoly_T_max}
   \left [\sum_{k=0}^{N_\omega} \omega_i \cdot \left(t^{{B}}_{{go}_i}\right)^{k} \right]^2{T^2_{\max}}  = \alpha^2  \left[ \Vert{\bar{\bf Z}_{r}(t_f)}\Vert^2\cdot \left(t^{B}_{{go}_i} \right)^2 + 2\bar{\bf Z}^T_{r}(t_f) \bar{\bf Z}_{V}(t_f)\cdot t^{B}_{{go}_i}  + \Vert{\bar{\bf Z}_{V}(t_f)}\Vert^2 \right]  \quad \forall t^{B}_{{go}_i} 
   ,\quad  i \in \left[ 1,\lvert t^{B}_{go} \rvert \right]
 \end{equation} 
\end{subequations}
For given values of $ \bar{\mathbf{Z}}_{r}(t_f)$ and $ \bar{\mathbf{Z}}_{V}(t_f) $, the switching times are obtained as the roots of \eqref{eqn:2ndorderpoly_T_max}. Furthermore, for a given value of $t^{B}_{go}$, the vector  $\bar{\mathbf{Z}}_{xy} (t_f)$ is obtained by solving the nonlinear system in \eqref{eqns:NonInvertabelZtfToZt_fpoly2nd}. The solution procedure for this system is presented in \secref{APP:IterativeScheme}.


\section{Implementation}
\label{APP:IterativeScheme}

\subsection{Algorithm Flow and Mass Approximation}
\label{5.1_MassApproximation}
\begin{figure} [ht] \centering 
\begin{center}
\begin{tikzpicture}[node distance=0.35cm and 1.25cm]
\node (start) [startstop] {Start};
\node (unconstrained) [process, below=of start] {OPDG ${{\bf Z}_f}$ };
\node (getf)  [process, below=of unconstrained]  {$f(t_{go}), h_{\max}(t_{go})$};
\node (OBWPDG) [process, below=of getf]  {Open-loop OBPDG $\bar{\bf Z}_{f}$   };
\node (ftgo) [process, below=of OBWPDG] {$f(t_{go}), t_{go}^{B},h_{\max}(t_{go})$};
\node (check) [decision, below=of ftgo] {Has $f(t_{go})$ converged?};
\node (tis0) [process, right=of check]  {$t=0$};
\node (OBWPDG_Closed) [process, below=of tis0]  {Open-loop OBPDG $\bar{\bf Z}_{f}$   };
\node (update_state) [process,below=of OBWPDG_Closed]  {Update $t,{\bf x}(t),f, h_{\max}$} ;
\node (tfort) [decision, left=of update_state] {$t\leq t_f$};
\node (end) [startstop, below=of tfort] {End};

\draw [arrow] (start) -- (unconstrained);
\draw [arrow] (unconstrained) -- (getf);
\draw [arrow] (getf) -- (OBWPDG);
\draw [arrow] (OBWPDG) -- (ftgo);
\draw [arrow] (ftgo) -- (check);
\draw [arrow] (check) -- node[above] {Yes} (tis0);
\draw [arrow] (tis0) -- (OBWPDG_Closed);
\draw [arrow] (OBWPDG_Closed) -- (update_state);
\draw [arrow] (update_state) -- (tfort);
\draw [arrow] (tfort.north)  -- ++(0,0.42) -- node[above] {Yes} (OBWPDG_Closed.west) ;
\draw [arrow] (tfort.south) -- node[left] {No} (end);
\draw [arrow] (check.west) -- ++(-1.5,0) node[left] {No} -- ++(0,2.87) -- (OBWPDG.west);
\end{tikzpicture}
\end{center}
\caption{Iterative solution scheme for the bounded problem.}
\label{BD:process}
\end{figure}
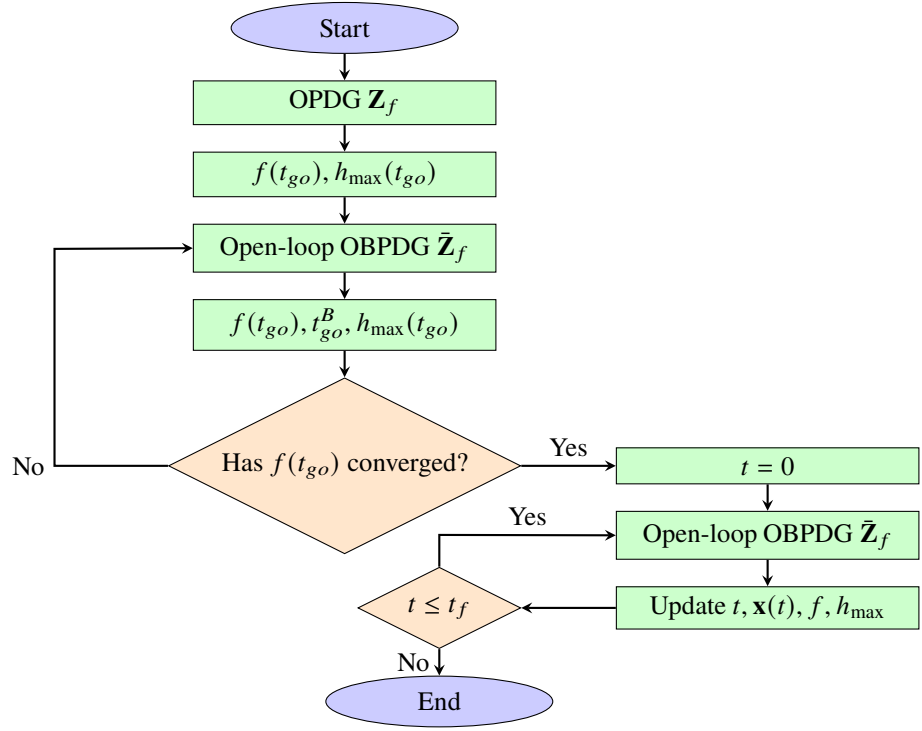

The first step is to solve a simplified unconstrained version of the problem. This problem admits an analytical solution, which is used to approximate the lander's mass. The resulting guidance law, referred to as Optimal Powered Descent Guidance (OPDG), is based on the analytical solution of \cite{Gutman} and is summarized in Appendix B.
Substituting the mass estimate into $f(t_{go})$ yields a function that is subsequently approximated by a second-order polynomial.
The vertical thrust-acceleration profile is left unconstrained, while the horizontal thrust-acceleration profile is subject to a hard constraint determined by the available thrust margin.
Therefore, the function $h_{\max}(t_{go})$ is computed and approximated by a second-order polynomial.
The second step aims to improve the mass approximation by applying the full algorithm with the thrust bounds, while using the second-order polynomial approximations of $f(t_{go})$ and $h_{\max}(t_{go})$ obtained in the first step.
This procedure is repeated until $f(t_{go})$ converges. 
The final step implements a pseudo closed-loop guidance scheme by repeatedly recomputing the open-loop Optimal Bounded Powered Descent Guidance (OBPDG) solution proposed in \secref{sec:OWBPDG_GuidanceLawDerivation} throughout the trajectory.
In each guidance update, the approximations of $f(t_{go})$ and $h_{\max}(t_{go})$ are updated using the mass computed during the previous update, while the initial guess for $\bar{\bf Z}(t_f)$ is taken from the corresponding solution obtained at that update.
A flowchart illustrating this procedure is shown in \figref{BD:process}.
%
\subsubsection{Mass Evaluation using the Thrust-Acceleration} 
The mass rate of the lander is opposite in sign to the fuel consumption rate $\dot{m}_f$; therefore
\begin{equation}  \label{eqn:rocketequation}
        \dot{m} = -\dot{m}_f = -\frac{\Vert\bf{T} \Vert}{I_{sp}g_0} 
\end{equation}
Substituting \eqref{v_dot}
into \eqref{eqn:rocketequation} yields
\begin{equation}
    \dot{m} =- \frac{m\Vert\bf{u} \Vert}{I_{sp}g_0} 
\end{equation}
and integrating the equation yields 
\begin{equation} \label{eqn:massintime} 
    m(t) =m_0 \cdot \exp \left( -\int ^{t}_{0}\frac{\Vert\bf{u} \Vert}{I_{sp}g_0}\,dt \right)  
\end{equation}

\subsubsection{Initial approximation using OPDG} \label{subsubsec:InitialApproximationOWPDG}
The first step of the algorithm employs OPDG to obtain an initial second-order approximation of $f(t_{go})$. The corresponding analytical solution is known, and the optimal control ${\bf u}^*$ is given by \eqref{eq:unbounded_optimal_controller}.
Substituting 
the optimal controller with ${\bf M}= \alpha{[{\boldsymbol I}]}_{6 \times 6}$ into  \eqref{eqn:massintime}, yields
\begin{equation} \label{eqn:masscons}
    m(t) = m_0 \cdot  exp \left \{ -\frac{\alpha}{g_0I_{sp}} \int ^{t}_{0}{\sqrt{\Vert{\bf Z}_{r}(t_f)\Vert^2 t^2_{go}+2{\bf Z}_r^T(t_f){\bf Z}_{V}(t_f) t_{go} + \Vert {\bf Z}_{V}(t_f) \Vert^2}} \,dt
    \right \} 
\end{equation}
Let
\begin{equation}
    \nu_1 = \Vert{\bf Z}_{r}(t_f)\Vert^2,  \quad \nu_2 = 2{\bf Z}_r^T(t_f){\bf Z}_{V}(t_f), \quad \nu_3=\Vert{\bf Z}_{V}(t_f)\Vert^2
\end{equation}
Substituting into \eqref{eqn:masscons} and integrating \cite{handbook}, yields
\begin{equation}
    m(t_{go}) = 
   m_0 \cdot exp \left \{ \frac{\alpha}{g_0I_{sp}}  
    \left [ {K \ln\left(\left|\zeta (t_{go})\right|\right) +
    \phi (t_{go})}
    - {K \ln\left(\left|\zeta (t_{go}=t_f)\right|\right) +
   \phi (t_{go}=t_f) }
    \right] \right \}
\end{equation}
where
\begin{equation} \label{eqns:qkg_def}
    K = 
    \frac{4\nu_1\nu_3-\nu_2^2}{8\nu_1 \sqrt{\nu_1}}, \quad
    \phi (t) =
    \frac{2\nu_1 t_{go}+\nu_2}{4\nu_1}, \quad
    \zeta (t) = 2\sqrt{\nu_1 }\sqrt{\nu_1 t_{go}^2+\nu_2 t_{go}+\nu_3}+2\nu_1 t_{go}+\nu_2 
\end{equation}
and therefore,
\begin{equation}
    f(t_{go}) = 
   \frac{1}{m_0} \cdot exp \left \{ \frac{-\alpha}{g_0I_{sp}}  
    \left [ {K \ln\left(\left|\zeta (t_{go})\right|\right) +
    \phi (t_{go})}
    - {K \ln\left(\left|\zeta (t_{go}=t_f)\right|\right) +
   \phi (t_{go}=t_f) }
    \right] \right \}
\end{equation}
Using a set of samples of $f(t_{go})$, a polynomial of arbitrary order $N_\omega$ is fitted via the least-squares (LS) method \cite{LeastSQ} to obtain the approximation $\hat{f}(t_{go})$. This approximation is then used to compute the switching times through \eqref{eqn:Stgo_generic} and to construct the approximation $\hat{h}_{\max}(t_{go})$. The same procedure is applied throughout the remainder of the algorithm, where updated approximations of $f(t_{go})$ and $h_{\max}(t_{go})$ are obtained from sampled function values using the LS method.

\subsubsection{Final Approximation using OBPDG}
The problem is then solved in open loop using the approximation $\hat{f}(t_{go})$ obtained in the previous step. To simplify the implementation, the controller generated by the full algorithm is integrated numerically, although analytical integration is also possible by partitioning the time domain into constrained and unconstrained intervals. The resulting functions $f(t_{go})$ and $h_{\max}(t_{go})$ are subsequently evaluated and re-approximated, and the procedure is repeated until a prescribed convergence criterion is satisfied.
The convergence criterion is based on the discrete $L_1$ norm of the difference between successive iterations. Specifically, $f(t_{go})$ is sampled at $N_{\mathrm{samples}}$ points, and convergence is declared when the sum of the absolute differences between iterations $k$ and $k+1$ at the sampled times falls below a prescribed threshold.
\begin{equation}
     \sum_i^{N_{samples}} \vert f[k+1](t_i)-f[k](t_i) \vert < e
\end{equation}

\subsubsection{Closed-loop Approximation using OBPDG}
A pseudo closed-loop solution is obtained by implementing a sequence of open-loop controllers. At each time step, the function $f(t_{go})$ is re-approximated using the current state and employed in the computation of the next control command.

\subsection{Calculating $\bar{\bf Z}_{xy}(t_f)$ and the switching times}
This section develops a numerical algorithm for solving the nonlinear system in \eqref{eqns:ZandST} under second-order polynomial approximations of $f(t_{go})$ and $h_{\max}(t_{go})$.

Let
\begin{equation}
    \textbf{F} = \begin{bmatrix}
        {F}_1 \quad 
        \textbf{F}^T_2
    \end{bmatrix}^T
\end{equation}
where
\begin{subequations}
    \begin{equation} \label{eqn:F1_def}
{F}_1 =    
        \left (\omega_0 +\omega_1 \cdot t^{{B}}_{{go}_i} +\omega_2 \cdot \left(t^{{B}}_{{go}_i}\right)^2 \right)^2 T_{\max}^2- \alpha^2  \left[ \Vert{\bar{\bf Z}_{r}(t_f)}\Vert^2\cdot \left(t^{{B}}_{{go}_i}\right)^2 + 2\bar{\bf Z}^T_{r}(t_f) \bar{\bf Z}_{V}(t_f)\cdot t^{B}_{{go}_i}  + \Vert{\bar{\bf Z}_{V}(t_f)}\Vert^2 \right]  \quad \forall t^{B}_{{go}_i} 
\end{equation} 
\begin{equation} \label{eqn:F2_def}
    \textbf{F}_2 =      \bar{\bf Z}_{xy}(t_f)- \bar{\bf Z}_{xy}(0) -
     \bar{\bf I}^{xy}_{NB} \left({\mathbb{R}_{NB}} \right) - 
     \bar{\bf I}^{xy}_{B}({\mathbb{R}_{B}}) 
\end{equation}
\end{subequations}
Thus, the roots-finding problem can be written as
\begin{subequations}  \label{eqn:rootfinding}
  \begin{align}
  \label{eqn:rootfinding_1}
    {F}_1 &= 0 \quad \forall t^{B}_{{go}_i} 
\\ \label{eqn:rootfinding_2}
    \textbf{F}_2 &= [\textbf{0}]_{4 \times 1}
  \end{align}
\end{subequations}
The proposed algorithm employs an iterative approach to solve the root-finding problem in \eqref{eqn:rootfinding}, assuming that $f(t_{go})$ and $\bar{\bf Z}_{z}$ are known. The algorithm consists of two stages: first, solving an equation, and second, using its solution to solve the remaining set of equations. The procedure is initialized with an estimate of $\bar{\bf Z}_{{xy}}(t_f)$ obtained from the OPDG solution. In a pseudo closed-loop implementation, the initial estimate is instead taken from the OBPDG solution computed at the previous guidance update.
For a given value of $\bar{\bf Z}_{xy}(t_f)$, the equation in \eqref{eqn:F1_def} reduces to a fourth-order polynomial, whose roots can be computed numerically using MATLAB's \texttt{roots} function. This function determines the polynomial roots by computing the eigenvalues of the associated companion matrix \cite{Matlab_root}.
Substituting these solutions into \eqref{eqn:F2_def} yields a system of four nonlinear equations with $\bar{\bf Z}_{xy}(t_f)$ as the unknown. Several numerical schemes may be used to solve this system. In the present work, the Newton-Raphson algorithm is adopted \cite{LeastSQ}.

The Following pseudo-code describes the Algorithm 
\begin{enumerate}
\item \label{step1}
{\textbf{Initial Guess:}}
    \begin{enumerate}[label=\roman*)]
     \item
    Compute $\bar{\bf Z}_{xy}(t_f)$ from the OPDG solution in Appendix B and use it as the initial guess. In a pseudo closed-loop implementation, the initial guess is instead taken from the OBPDG solution obtained at the previous guidance update.
      \item  Approximate $f(t_{go})$ using the procedure described in \secref{5.1_MassApproximation}.
    \item 
    Define $m=0$ as a counter of the steps to converge the solution for ${\bf F} = [{\bf 0}]_{5 \times 1}$.
    \item \label{st}
    Compute the switching times ${\bf t}^{B}_{go}[0]$ by solving \eqref{eqn:rootfinding_1} using the calculated value of $\bar{\bf Z}_{xy}(t_f)$. Equation \eqref{eqn:rootfinding_1} reduces to a fourth-order polynomial, which is solved numerically using MATLAB's \texttt{roots} function.
    \item 
    Approximate $h_{\max}(t_{go})$ using the procedure described in \secref{5.1_MassApproximation}.
    \end{enumerate}
\item \label{step2}
{\textbf{Iterative process:}}
\begin{enumerate}[label=\roman*)]
    \item 
    Define ${\mathbb{R}_{B}}$ 
    \eqref{eqns:conditions_for_time_intervals}.
    \item
    Substitute the switching times and $\bar{\bf Z}(t_f)[m]$  into the Jacobian matrix inverse.
    \begin{equation}
        \textbf{J}^{-1}_2(\bar{\bf Z}_{{xy}} (t_f))[m] =\left (\frac{\partial {\bf F}_2 }{\partial \bar{\bf Z}_{xy}(t_f)} \right) ^{-1}
    \end{equation}
    \item 
    Define $k=0$ as a counter of the steps to converge the solution for ${\bf F}_2 = [{\bf 0}]_{4 \times 1}$.
    \item
    Apply the Newton-Raphson algorithm to obtain the $\textbf{F}_2={\bf 0}$ solution.
        \begin{enumerate}[label=\Roman*)]
        \item \label{stepI}
        Update $\bar{\bf Z}_{{xy}}(t_f)$ using the scheme
        \begin{equation}
            \bar{\bf Z}_{xy}(t_f)[k]=\bar{\bf Z}_{xy}(t_f)[k-1] - {\bf J}_2^{-1} \cdot \textbf{F}_2 [k-1]
        \end{equation}
        \item 
        Calculate the error as
           $ e_k = \Vert\textbf{F}_2[k]\Vert$.
        \item 
        Update $k = k+1$. \label{stepIII}

        \item 
        Repeat steps \ref{stepI}-\ref{stepIII} until the error $e_k$ exceeds the desired convergence criteria. 
    \end{enumerate}
\end{enumerate}
\item 
\textit{\textbf{Conversion:}}
    \begin{enumerate}[label=\roman*)]
    \item \label{stepIV}
    Compute the switching times $\ {\bf t}^{B}_{go}[m]$ by solving \eqref{eqn:rootfinding_1} using the computed value of $\bar{\bf Z}_{xy}(t_f)$. The resulting fourth-order polynomial is solved numerically using MATLAB's \texttt{roots} function.
    \item 
    If the number of switching times differs from that obtained in the previous iteration, repeat the procedure.
    \item
    Calculate the error as
      $ e_m = \left\Vert 
      \left({\bf t}^{B}_{go}[m]- {\bf t}_{go}^{B}[m-1]\right)^T \quad  \textbf{F}^T_2 \right\Vert  $.
    \item
    Update $m = m+1$.
    \item 
    Repeat the procedure until the error $e_m$ falls below the prescribed convergence tolerance.
    \end{enumerate}
\end{enumerate}

\section{Performance Analysis} \label{sec:PerformanceAnalysis}
\subsection{Simulated Scenario}\label{Simulated Scenario}
The numerical simulations integrated the equations of motion in \eqref{eq:motion_eqn} using the optimal open-loop controller given by \eqref{eqn:foptimalcontrollercases} and second-order polynomial approximations of $f(t_{go})$ and $h_{\max}(t_{go})$. To represent a realistic implementation and improve robustness, a pseudo closed-loop strategy was adopted, in which the open-loop optimal control problem was re-solved at a rate of 1 Hz throughout the trajectory.
The equations of motion were integrated using MATLAB's \texttt{ode45} solver, which implements an explicit Runge-Kutta method. The simulations were performed using the initial conditions and parameters of \cite{1997} with slight changes to demonstrate the proposed algorithm.
The initial conditions and problem parameters are summarized in Table \ref{tbl:IC}. The final time was selected as the largest admissible value that avoids ground collision under the OPDG solution, as determined from \eqref{eq:t_max}. For the considered initial conditions, the maximum admissible final time is $t_f=304.8 \, (s)$, and therefore $t_f=304 \, (s)$ was used in the simulations. 

For the same initial conditions, the unconstrained optimal final time predicted by the OPDG solution is $t_f^*=428 \, (s)$ \cite{1997}. However, this value exceeds the maximum admissible final time required for ground-collision avoidance and is therefore infeasible.
The function $f(t_{go})$ was updated in each open-loop iteration.

\begin{figure} [htbp]
 \centering
    \includegraphics[scale  = 0.65]{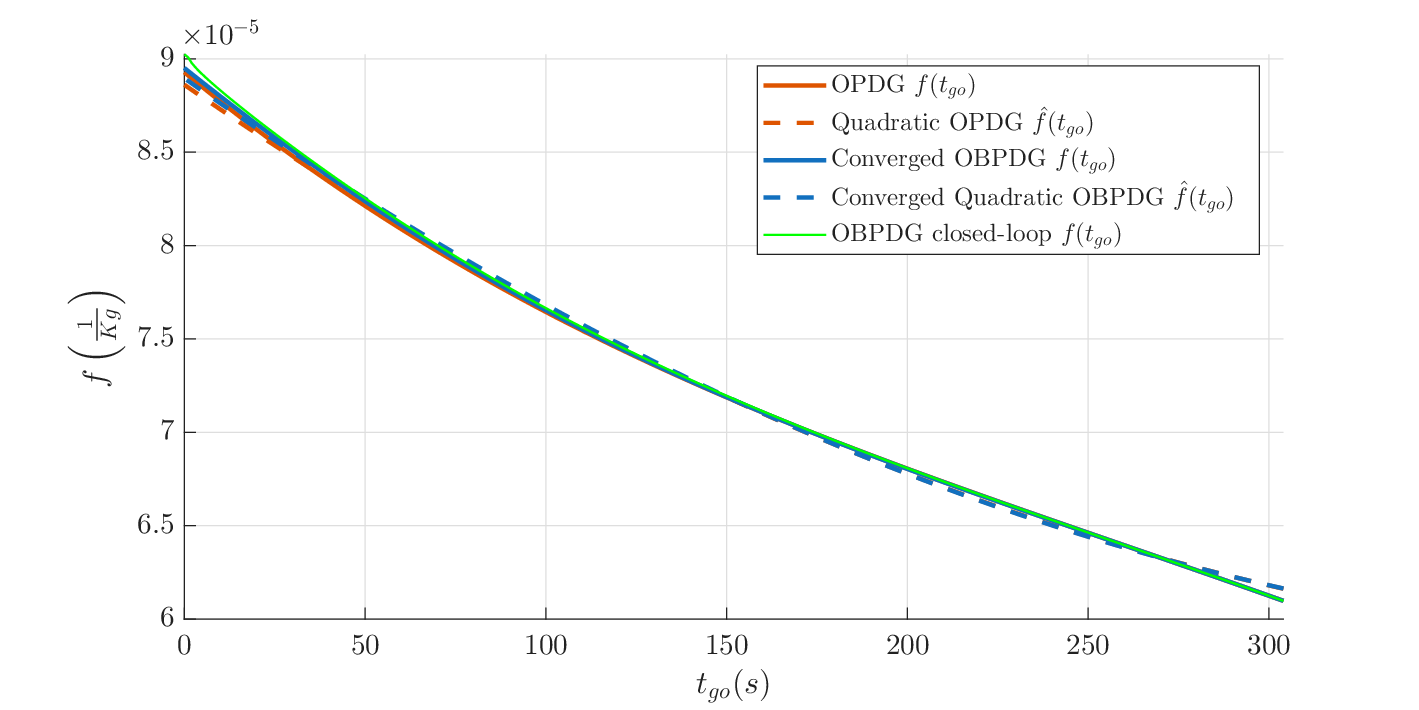}
    \caption{Approximation of $f(t_{go})$.}
    \label{fig:ftgo}
\end{figure}

\begin{table}[h]
    \centering
    \begin{tabular}{|c|c|c|}
        \hline
        \textbf{Parameter} & \textbf{Value}  & \textbf{Unit} \\
        \hline
       $\alpha$ & $10^{6}$ & - \\
       $ {\bf r}(0)$ & $[152400, 30480,-15240]^T $ & $m$ \\ 
       ${\bf V}(0)$ & $[-800,0,150]^T$ & $\frac{m}{s}$\\ 
       ${\bf g}$ & $[0,0,1.615]^T$ & $\frac{m}{s^2}$\\
        $t_f$ & $304$ & $s$ \\
        $T_{\max}$ & $56$ & $KN$ \\ 
        $I_{sp}$ & $300$ & $s$ \\
        $m(0)$ & $16400$ & $Kg$ \\ 
        \hline
    \end{tabular}
    \caption{The initial conditions and parameters of the problem. }
    \label{tbl:IC}
\end{table}
Figure \ref{fig:ftgo} shows $f(t_{go})$ and its second-order polynomial approximations at various stages of the initialization process. The initial approximation is obtained from the OPDG solution. Subsequently, the algorithm switches to the OBPDG solution and iteratively refines the approximation of $f(t_{go})$ until convergence is achieved at $t=0$. The resulting converged function and its second-order polynomial approximation are presented.
Note that, because the solution is implemented as a sequence of open-loop controllers, the approximation of $f(t_{go})$ is recomputed at each guidance update to improve accuracy and update the controller bounds. The resulting pseudo closed-loop approximations are therefore also included in the figure. Since each update is initialized using the approximation obtained at the previous guidance update, the full initialization procedure is not required.

The coefficients of the polynomial approximations of $f(t_{go})$ at the various stages of the algorithm are presented in Table \ref{tbl:Coeffs_2ndOrder}. The similarity of the resulting approximations indicates that the OPDG provides sufficiently accurate initial estimates, thereby justifying its use for initialization and for generating the initial approximations of $f(t_{go})$ and $h_{\max}(t_{go})$.
\begin{table}[h] 
    \centering
    \begin{tabular}{|c||c|c|c|} 
        \hline
        \textbf{Outline} & $a_2 \cdot 1e-10$ & $a_1 \cdot 1e-7$ & $a_0 \cdot 1e-5 $ \\ 
        \hline
        OPDG &  1.476                                & -1.336  &  8.86      \\
        Converged initialization  & 1.544  &  -1.368   &  8.896  \\
        Closed-loop OBPDG & 1.596  &  -1.393   &   8.927 \\
        \hline
    \end{tabular}
    \caption{$f(t_{go})$ coefficients in different initiation steps of the algorithm.}
    \label{tbl:Coeffs_2ndOrder}
\end{table}
 
\subsection{Sample run}
\begin{figure} [htbp]
 \centering
    \includegraphics[scale  = 0.65]{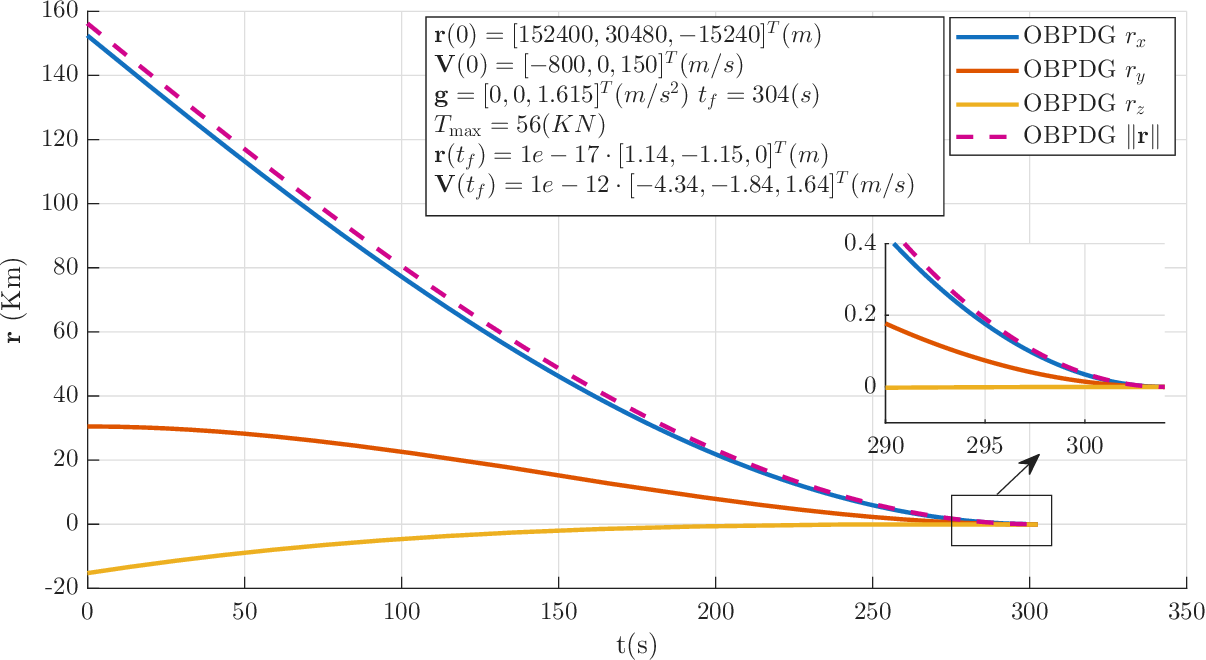}
    \caption{Sample Run - Position time history. }
     \label{fig:rallaxis}
\end{figure}

\begin{figure} [htbp]
 \centering
    \includegraphics[scale  = 0.65]{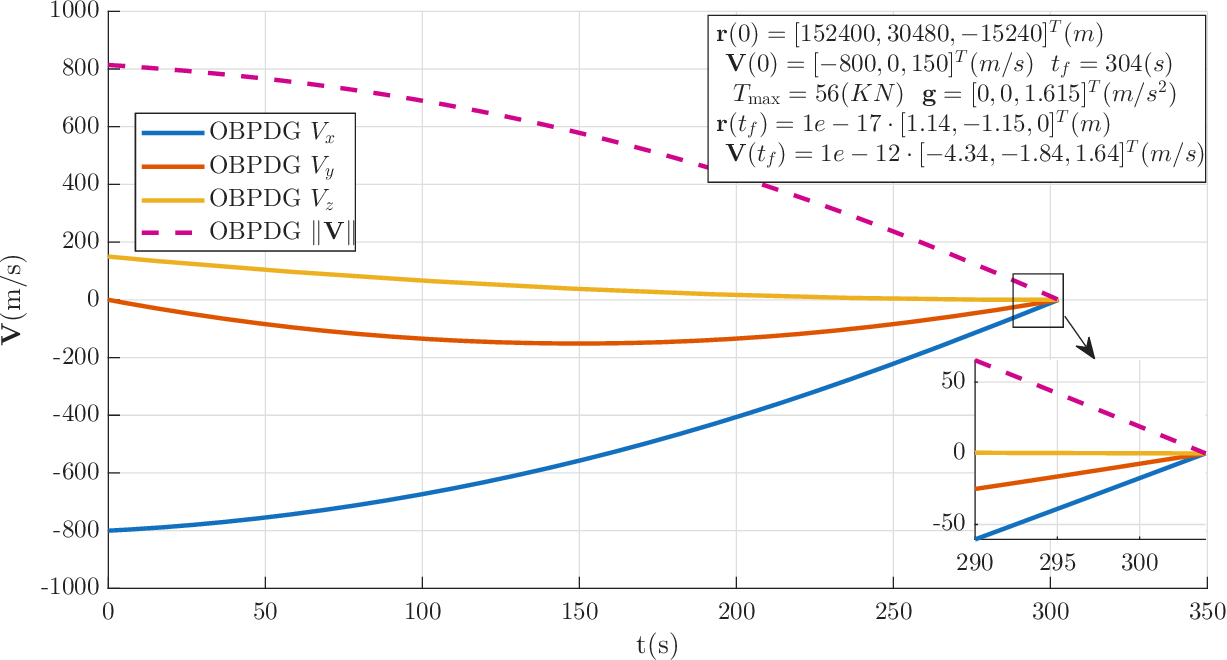}
    \caption{Sample run - Velocity time history.}
    \label{fig:vallaxis}
\end{figure}

This section demonstrates the performance of the proposed OBPDG scheme in a representative simulation. The position and velocity time histories are shown in \figref{fig:rallaxis} and \figref{fig:vallaxis}, respectively. The lander successfully performs a soft landing at the designated landing site, satisfying the terminal position and velocity constraints to numerical precision. The terminal position and velocity errors are on the order of $10^{-17} \, (m)$ and $10^{-12} \, (m/s)$, respectively.

\begin{figure} [htbp]
 \centering
    \includegraphics[scale  = 0.65]{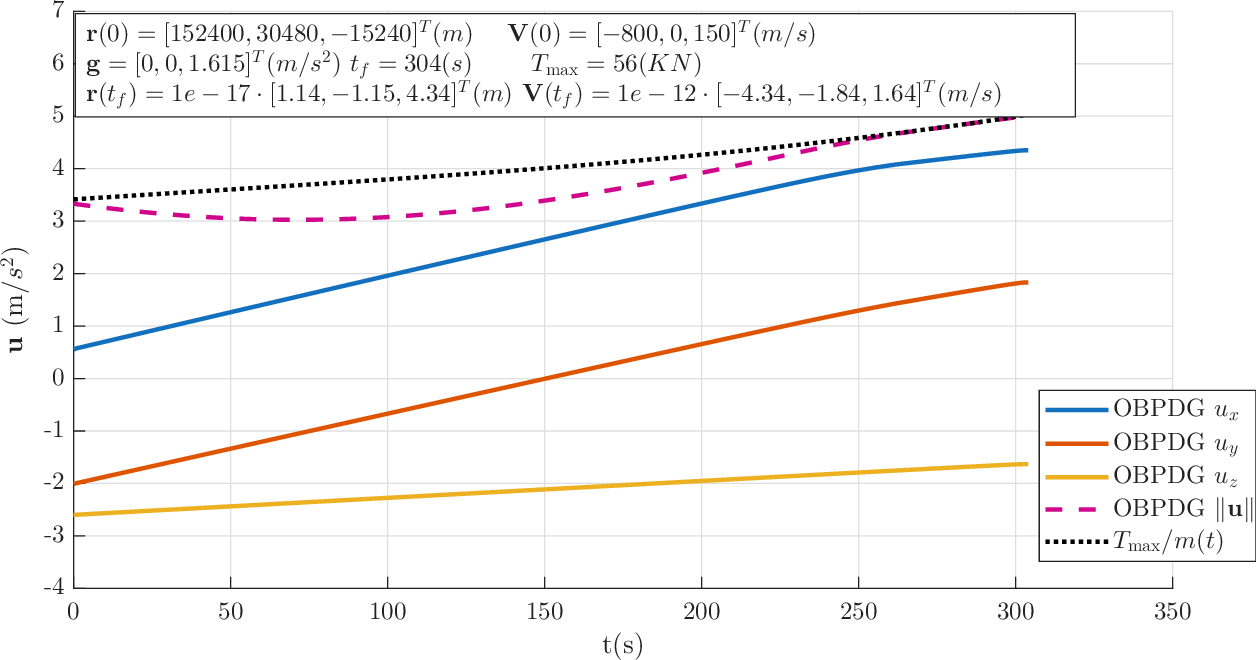}
    \caption{Sample Run - Thrust-acceleration time history.}
     \label{fig:uallaxis}
\end{figure}

The thrust-acceleration components and their magnitude are shown in \figref{fig:uallaxis}. The corresponding thrust-acceleration bounds are computed using \eqref{v_dot} and a numerical integration of the mass dynamics in \eqref{u_from_T}. The controller is saturated over a significant portion of the trajectory, particularly during the terminal phase of the descent. Despite this extensive saturation, the guidance law achieves a highly accurate soft landing by predicting the entry and exit times of the saturated region and compensating for the resulting loss of control authority in advance.

\subsection{Comparison between OPDG and OBPDG}
\begin{figure} [htbp]
 \centering
    \includegraphics[scale  = 0.65]{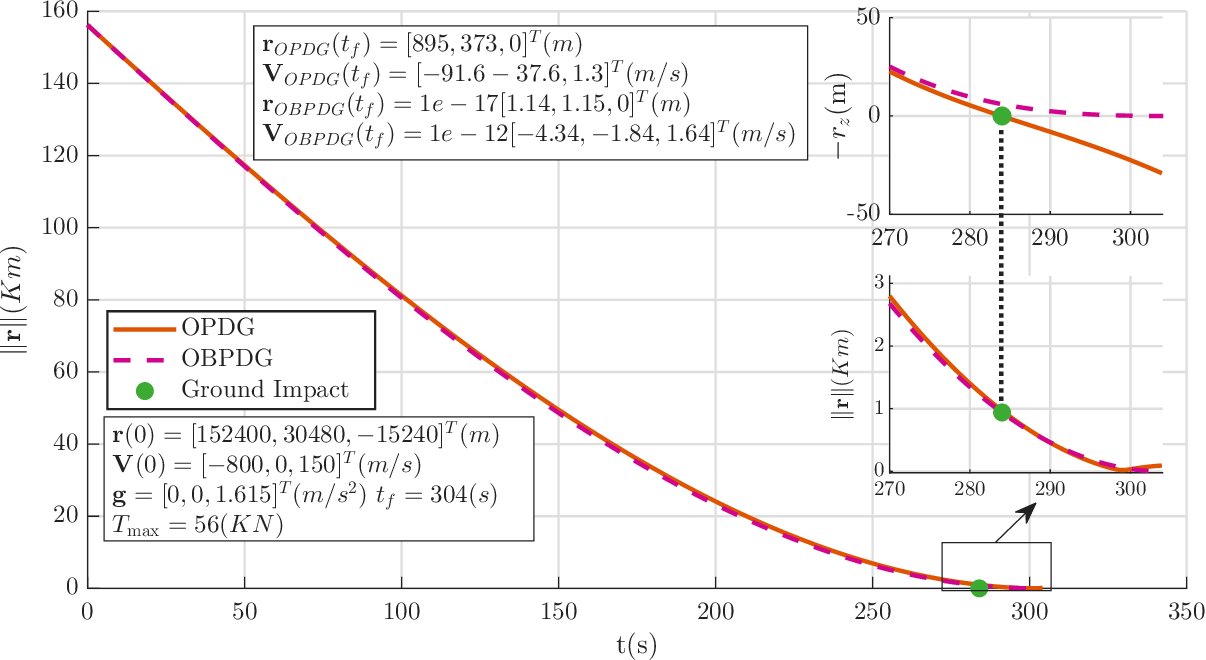}
    \caption{Comparison - Position time history. }
     \label{fig:rallaxis_cmp}
\end{figure}

\begin{figure} [htbp]
 \centering
    \includegraphics[scale  = 0.65]{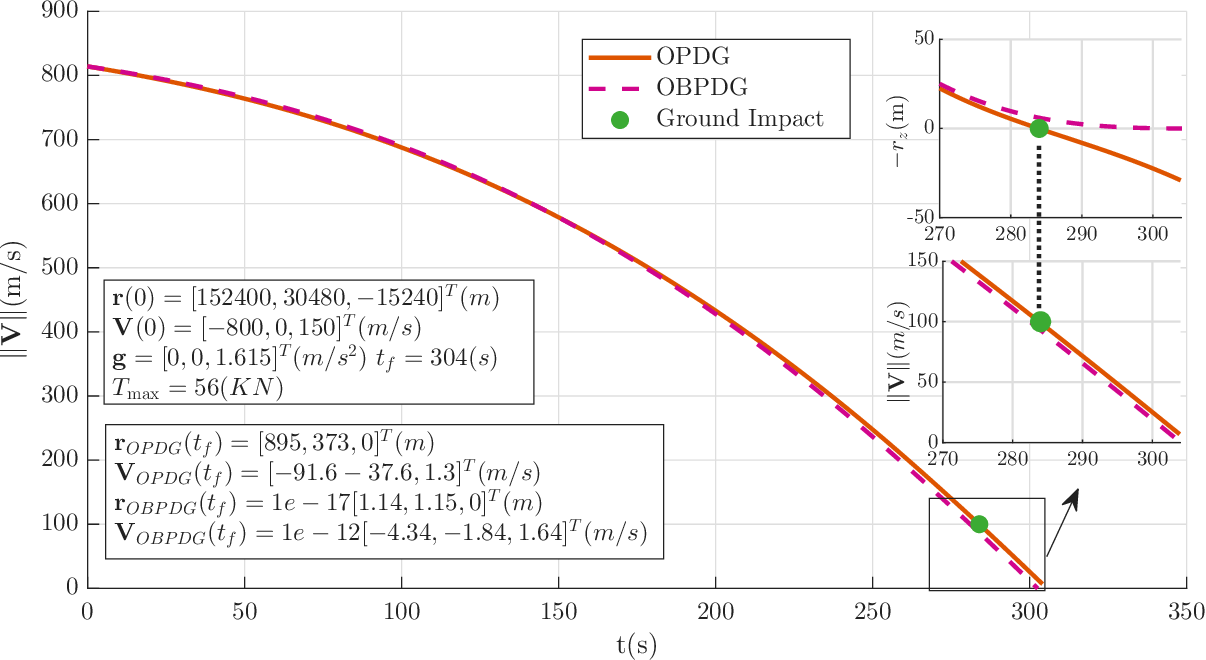}
    \caption{Comparison - Velocity time history.}
    \label{fig:vallaxis_cmp}
\end{figure}

This section compares the performance of OPDG and OBPDG. Figures \ref{fig:rallaxis_cmp} and \ref{fig:vallaxis_cmp} compare the position and velocity magnitudes obtained with the two guidance laws. While the lander guided by OBPDG successfully performs a soft landing at the designated landing site, the lander guided by OPDG collides with the ground approximately $20 \,(s)$ before the selected final time. At the point of impact, its velocity magnitude is $\Vert{\bf V}\Vert = 99 \, (m/s)$ and its miss distance is $\Vert{\bf r}\Vert = 970 \, (m)$. Note that the scenario duration was deliberately chosen to be slightly below the maximum final time permitted by the OPDG ground-collision-avoidance condition. Furthermore, due to the unanticipated saturation near the end of the maneuver, the velocity magnitude at the selected final time remains $\Vert{\bf V}(t_f)\Vert = 7 \, (m/s)$. Consequently, OPDG fails to achieve a soft landing even when the ground-collision constraint is ignored.
Table \ref{tab:Results_Overview} presents the terminal position, velocity, and the total impulse of the two guidance laws. 

\begin{figure} [htbp]
 \centering
    \includegraphics[scale  = 0.65]{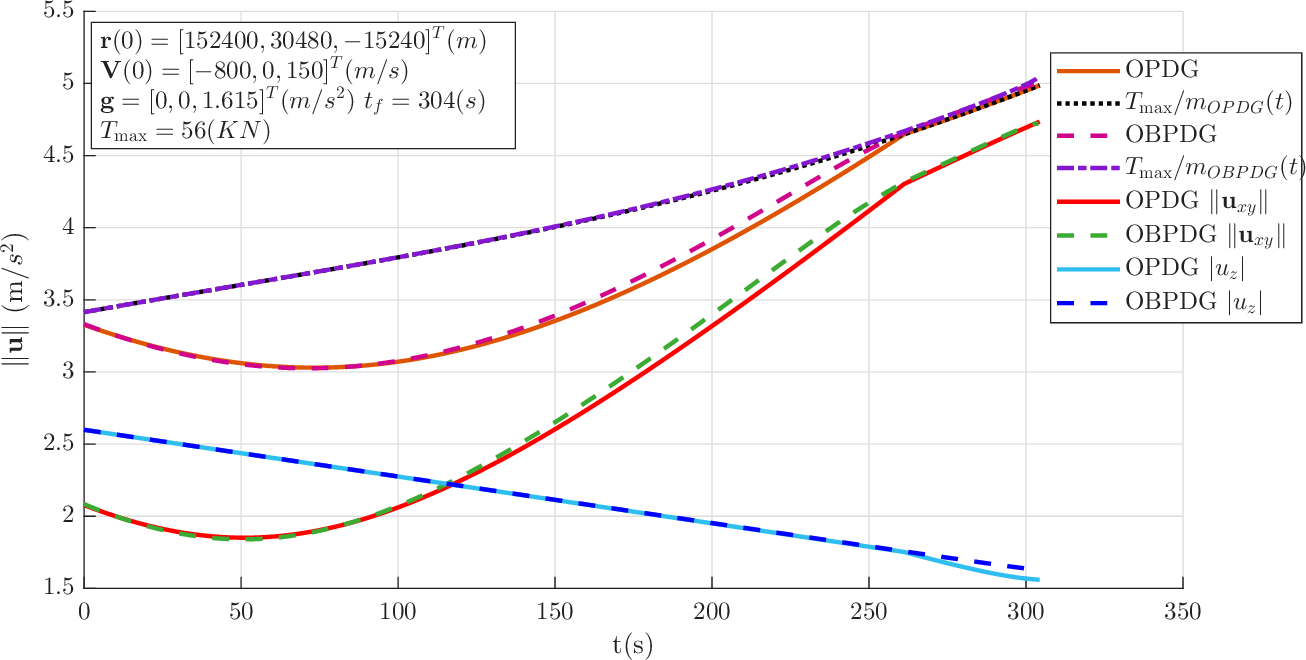}
    \caption{Comparison - Thrust-acceleration magnitude.}
    \label{fig:hut_cmp}
\end{figure}

\begin{figure} [htbp]
 \hspace{2.7cm}
    \includegraphics[scale  = 0.65]{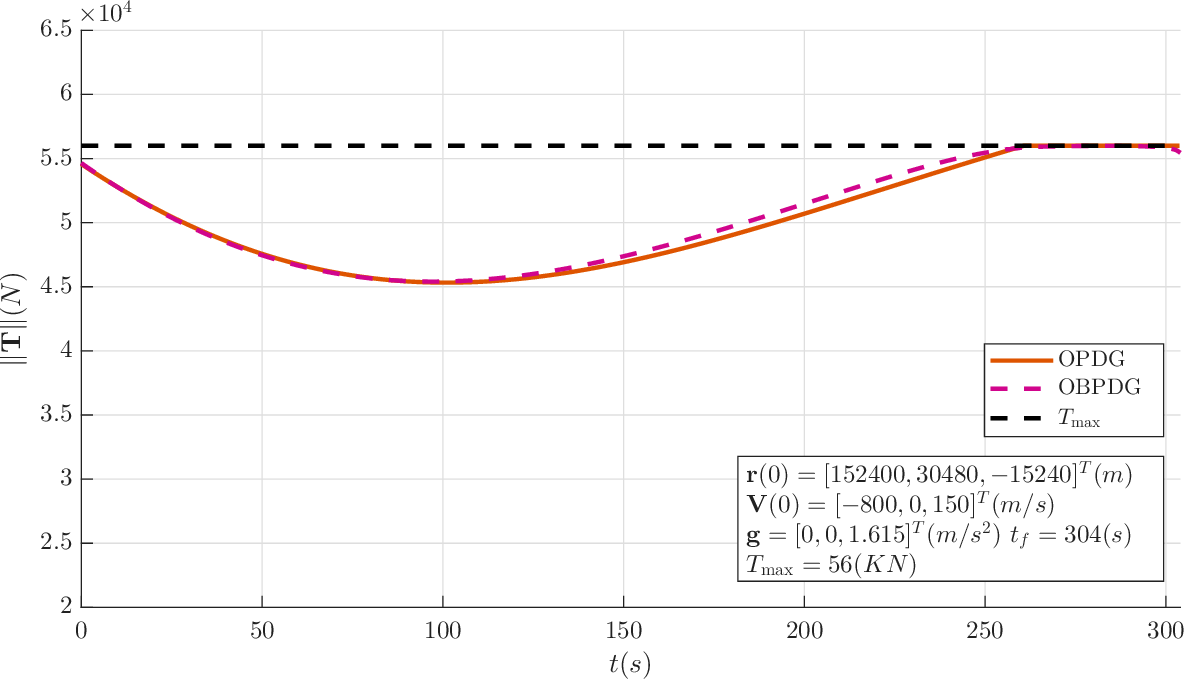}
    \caption{Comparison - Thrust magnitude.}
    \label{fig:T_t_cmp}
\end{figure}

Figures \ref{fig:hut_cmp} and \ref{fig:T_t_cmp} present the thrust-acceleration magnitude and thrust magnitude, respectively. Since OPDG does not account for the anticipated saturation, it fails to compensate for the resulting loss of control authority, leading to a target miss and ultimately to ground collision. In contrast, OBPDG predicts the onset of saturation and shapes the acceleration profile accordingly, thereby achieving a highly accurate soft landing despite the prolonged saturation.

This is accomplished by increasing the horizontal acceleration while maintaining the vertical acceleration profile prior to the predicted saturation interval. As shown in \figref{fig:hut_cmp}, this behavior becomes apparent approximately $100 \,(s)$ before the onset of saturation. Consequently, less control effort is required in the horizontal plane during the saturated phase, allowing a larger fraction of the available thrust to be allocated to the vertical axis. This additional vertical control authority reduces the accumulated trajectory errors during saturation and enables the lander to satisfy the terminal soft-landing conditions.

\begin{table}[htbp]
    \centering
    \begin{tabular}{|c|c|c|c|} \hline
      & $ {\bf r}(t_f)(m)$ & ${\bf V}(t_f)(m/s)$ & $ \int{\Vert {\bf T} \Vert} dt (N\cdot s)$  
      \\ \hline  
      OPDG   & $  [895,373,0]^T$ & $[-91.6,-37.6, 1.3]^T$       & $15.22 \cdot 10^6$   \\ 
\hline
      OBPDG & $1e-17[1.14,1.15,0]^T$ & $1e-12[-4.34,-1.84,1.64]^T$ & $   15.3 \cdot 10^6  $\\ \hline 
    \end{tabular}
    \caption{Performance comparison.}
    \label{tab:Results_Overview}
\end{table}


\input{Spaghetti.tex}

\section{Conclusions} \label{sec:Conclusions}
In this paper, a new optimal guidance law for the bounded-thrust powered-descent problem was developed. The lander mass was approximated using a time-dependent polynomial, yielding a bounded linear-quadratic optimal-control formulation that minimizes thrust-acceleration control effort while satisfying terminal position and velocity requirements. The mass approximation enables explicit incorporation of the mass-dependent thrust-acceleration limits induced by propellant consumption. A hierarchical thrust-allocation strategy was introduced, in which the vertical thrust profile remains unconstrained while the horizontal thrust profile is subject to a hard constraint determined by the available thrust margin. This formulation enables analytical prediction and avoidance of ground collision while explicitly accounting for thrust saturation.
The resulting guidance law predicts the entry into and exit from saturated regions and shapes the thrust-acceleration profile accordingly, allowing the lander to compensate for future loss of control authority. The proposed formulation retains much of the computational efficiency and analytical structure of classical powered-descent guidance laws while extending them to bounded-thrust scenarios with ground-collision avoidance. Consequently, accurate soft-landing performance is achieved even when the controller remains saturated for substantial portions of the trajectory. Extensive numerical simulations demonstrated robust performance over a grid of perturbed initial conditions and highlighted the significant degradation that can occur when thrust saturation is neglected.
By combining saturation-aware guidance with analytical ground-collision prediction and avoidance, the proposed method provides a computationally efficient framework for reliable collision-free powered descent under bounded thrust.

\section*{Appendix A: Bounded Integral with a Second Order Polynomial Approximations}
\input{Appendix_FirstOrderPolynomialApproximation_scitech}

\section*{Appendix B: Optimal Powered Descent Guidance}
In this Appendix, the OPDG guidance law derived in \cite{Gutman} is presented for completeness, as it forms the basis for several developments in the current paper. Substituting \eqref{eq:unbounded_optimal_controller} into \eqref{z_dot_with_b_tilde}
integrating from $t$ to $t_f$ and rearranging yields
 \begin{equation} \label{eqn:OWPDG_Ztf}
    \begin{bmatrix}
        {\bf Z}_{r}(t_f) \\ {\bf Z}_{V}(t_f) 
    \end{bmatrix}= 
 \begin{bmatrix}
   \boldsymbol{\Delta}^{-1}
     &
     {[{\boldsymbol 0}]}_{3 \times 3}
     \\ 
   {[{\boldsymbol 0}]}_{3 \times 3}
     &
     \boldsymbol{\Delta}^{-1}
 \end{bmatrix}
    \begin{bmatrix}
    {[{\boldsymbol I}]}_{3 \times 3} + t_{go} {\bf M}_{VV}^2
     &
     -\frac{1}{2}t^2_{go} {\bf M}_{rr} {\bf M}_{VV}
     \\ 
    - \frac{1}{2}t^2_{go} {\bf M}_{rr} {\bf M}_{VV}
     &
      {[{\boldsymbol I}]}_{3 \times 3} + \frac{1}{3}t^3_{go} {\bf M}_{rr}^2
 \end{bmatrix}
 \begin{bmatrix} 
        {\bf Z}_r(t) \\ {\bf Z}_V(t)
    \end{bmatrix}
 \end{equation}
\begin{equation} \label{eqn:Deltadef}
 \boldsymbol{\Delta}=\left[{[{\boldsymbol I}]}_{3 \times 3} + \frac{1}{3}t^3_{go} {\bf M}_{rr}^2 \right] \left[{[{\boldsymbol I}]}_{3 \times 3} + t_{go} {\bf M}_{VV}^2\right]- \frac{1}{4}t^4_{go} {\bf M}_{rr}^2 {\bf M}_{VV}^2
 \end{equation}
 where $\boldsymbol{\Delta}$ is a diagonal Positive-Definite (PD) matrix for all $t_{go}\ge 0$, and is therefore invertible. Its inverse is obtained by inverting the diagonal elements.
 The optimal closed-loop controller for the unbounded problem is obtained by substituting \eqref{eqn:OWPDG_Ztf} into \eqref{eq:unbounded_optimal_controller}

\input{OptimalTimeForOPDG}

\section*{Acknowledgment}
The authors acknowledge the use of ChatGPT and Grammarly for editorial assistance during manuscript preparation, including grammar correction and language refinement.

\bibliography{sample.bib}

\end{document}

%% file: command.tex
\newcommand{\bfphi}{\boldsymbol{\phi}}

\renewcommand{\eqref}[1]{Eq.~(\ref{#1})}

\newcommand{\figref}[1]{Fig.~\ref{#1}}

\newcommand{\secref}[1]{Sec.~\ref{#1}}

%% file: Spaghetti.tex
\subsection{Robustness to initial conditions}
\begin{figure}[htbp]
    \centering
    \includegraphics[scale  = 0.65]{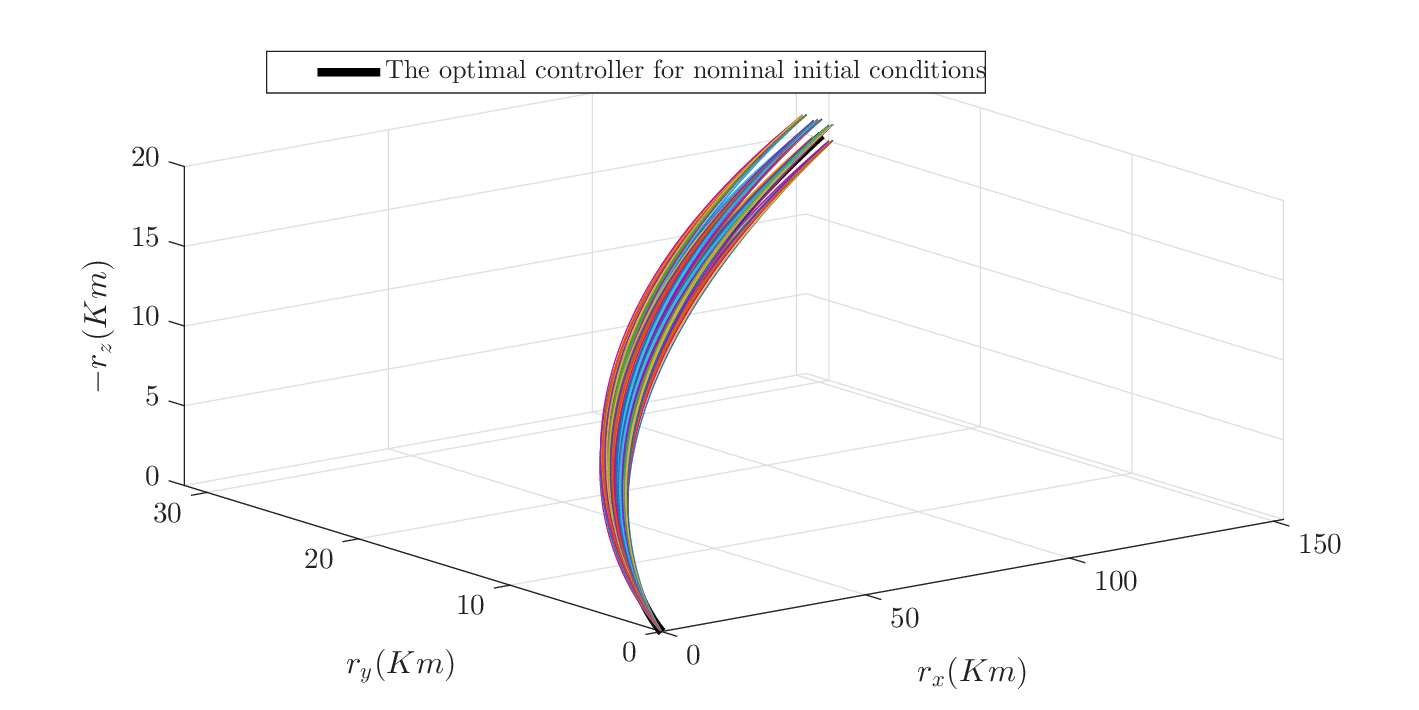}
    \caption{Robustness to initial conditions - Trajectories.}
    \label{fig:Robustness}
\end{figure}

Figure \ref{fig:Robustness} presents three-dimensional plots of the optimal trajectories obtained for a range of initial conditions to assess the robustness of the proposed algorithm. The nominal initial conditions were deliberately selected to be highly challenging and to induce long periods of control saturation, which is the primary operating regime targeted by the guidance law.
Specifically, the nominal initial position and velocity were set to $[151400,30480,-15240] \,(m)$ and $[-810,0,150] \,(m/s)$, respectively. A grid of test cases was generated by varying each position component within $\pm1000 \,(m)$ and each velocity component within $\pm10 \,(m/s)$ about the nominal values. The nominal scenario from the previous sections is included in the test set.
The perturbations in the $z$ direction were restricted to satisfy the condition in \eqref{eq:t_max} and thereby guarantee ground-collision avoidance.

\begin{figure}[htbp]
      \includegraphics[scale  = 0.65]{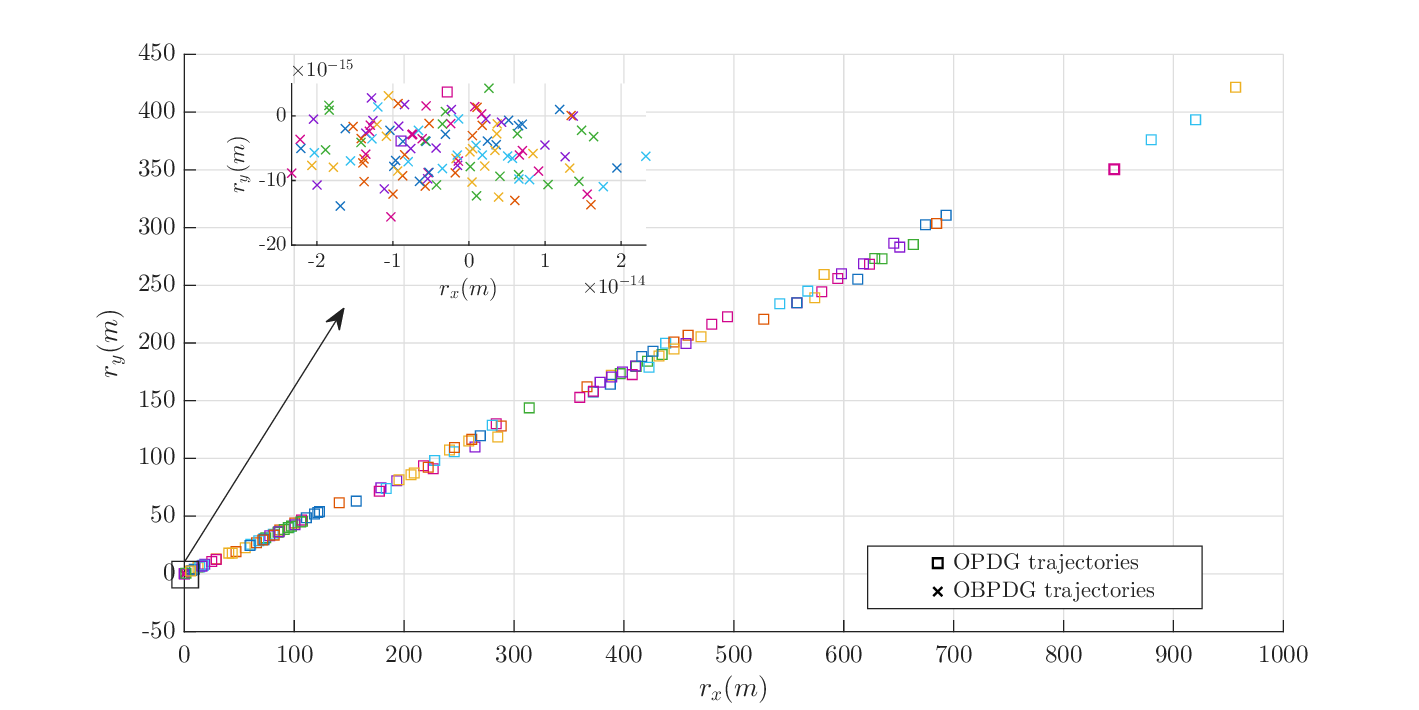}
      \caption{Robustness to initial conditions - Miss distances for OPDG and OBPDG.}
      \label{fig:rhits}
  \end{figure}

\begin{figure}[htbp]
  \includegraphics[scale  = 0.65]{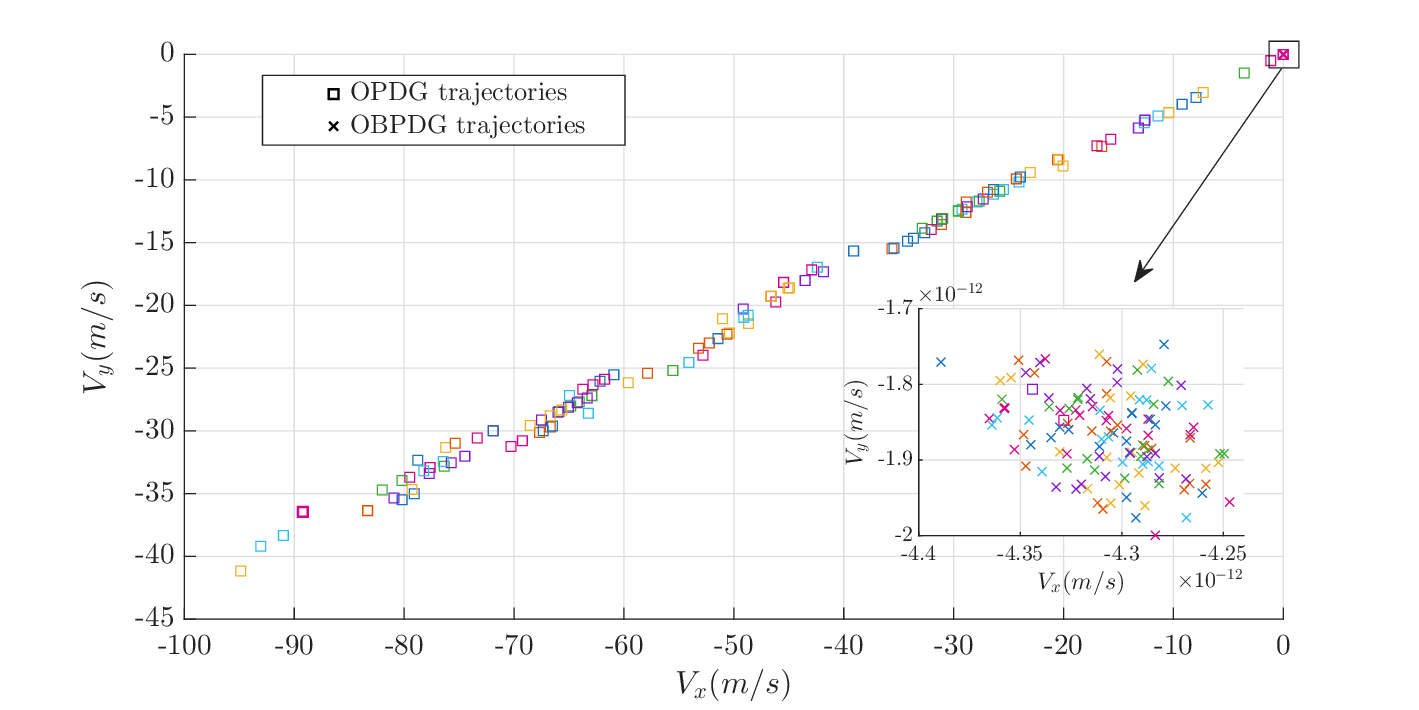}
      \caption{Robustness to initial conditions - Terminal velocities for OPDG and OBPDG.}
      \label{fig:vhits}
  \end{figure}

\begin{figure}[htbp]
\includegraphics[scale  = 0.65]{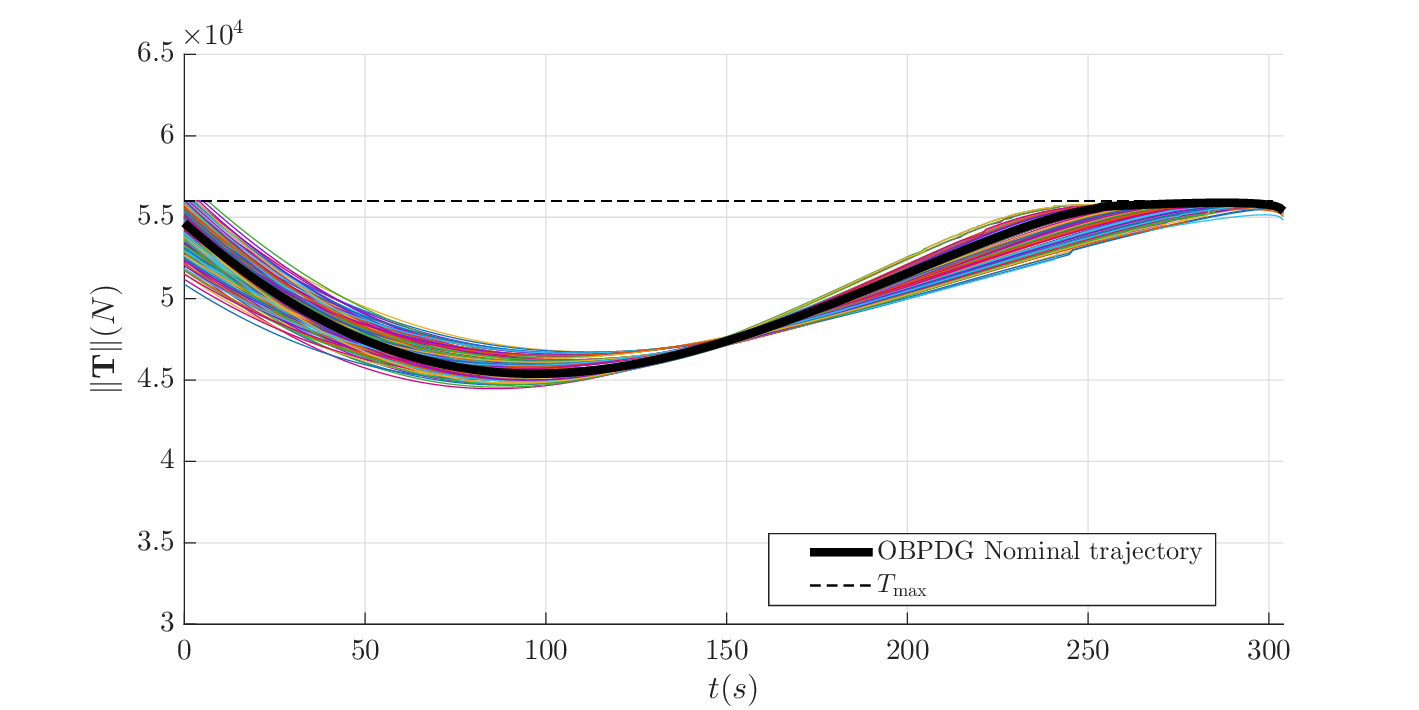}
    \caption{Robustness to initial conditions - Thrust profiles.}
    \label{fig:T_t_p}
\end{figure}

Figures \ref{fig:rhits} and \ref{fig:vhits} present the terminal position and velocity errors in the $x$-$y$ plane for OPDG and OBPDG. Owing to its ability to predict the saturated regions, OBPDG achieves negligible terminal position errors below $10^{-14} \, (m)$, whereas OPDG exhibits miss distances of up to $1000 \,(m)$.
Similarly, OPDG often reaches the terminal point with velocity errors as high as $100 \, (m/s)$, while OBPDG consistently achieves a soft landing at the target location with terminal velocity errors below $10^{-11} \, (m/s)$. The corresponding thrust profiles are shown in \figref{fig:T_t_p}. It is evident that all trajectories generated under the perturbed initial conditions satisfy the thrust bound.

%% file: Appendix_FirstOrderPolynomialApproximation_scitech.tex
\label{subsec:SOAPPROX}
In this Appendix, the analytical solution of the bounded integral $\bar{\bf I}^{xy}_{B}$ with a second-order polynomial approximation of $h_{\max}(t_{go})$ is derived. Let us assume a second-order polynomial approximation of $h_{\max}(t_{go})$
\begin{equation}
\label{eqn:fasSOpoly}
    h_{\max}(t_{go}) = a_0 + a_1 \cdot t_{go} +a_2 \cdot t^2_{go}
\end{equation}
To simplify the notations and due to the long expressions, we use abbreviated notations for this appendix only and denote 
\begin{equation}
    \bar{\bf Z}_{r_f} = \bar{\bf Z}_{{r}_{xy}} (t_f), \qquad \bar{\bf Z}_{V_f} = \bar{\bf Z}_{{V}_{xy}}(t_f) , \qquad \bar{\bf I}_{B}=\bar{\bf I}^{xy}_{B}
\end{equation}
Therefore, the indefinite integral associated with \eqref{eqn:fboundedintegral} reduces to
\begin{equation}
\label{eqn:boundedintegralFO}
  \begin{split}
    \bar{\bf I}_{B}
    & = -\alpha T_{\max}
     \int    \frac{
    a_0+a_1 \cdot t_{go} + a_2 \cdot t^2_{go}}{\sqrt{\Vert \bar{\bf Z}_{r_f}\Vert^2 t^2_{go}+2 \bar{\bf Z}_{r_f}^T \bar{\bf Z}_{V_f} t_{go} + \Vert \bar{\bf Z}_{V_f} \Vert^2}} \begin{bmatrix}
        t^2_{go} \bar{\bf Z}_{r_f}+ t_{go} \bar{\bf Z}_{V_f} \\ 
        t_{go} \bar{\bf Z}_{r_f} + \bar{\bf Z}_{V_f}
    \end{bmatrix} \,dt_{go}  
\end{split}  
\end{equation}
And the solution is
\begin{subequations}
\label{eqn:solutiontoIBwith2nd}
\begin{equation}
\begin{split}
        \bar{\bf I}_{B}(1:3)  
        & = \frac{-\alpha T_{\max}{\boldsymbol{ \mathcal K}}_1}{24  {\Vert \bar{\bf Z}_{r_f} \Vert}^9} \cdot \ln \left( \left|{\Vert \bar{\bf Z}_{r_f} \Vert}  \sqrt{{\Vert \bar{\bf Z}_{r_f} \Vert}^2  {t^2_{go}} + 2  { \bar{\bf Z}^T_{r_f} \bar{\bf Z}_{V_f} }   {t_{go}} + {\Vert \bar{\bf Z}_{V_f} \Vert}^2} + {\Vert \bar{\bf Z}_{r_f} \Vert}^2  {t_{go}} + { \bar{\bf Z}^T_{r_f} \bar{\bf Z}_{V_f} } \right| \right ) 
\\
        &  + \frac{-\alpha T_{\max}{\boldsymbol{ \mathcal G}}_1(t_{go})}{24  {\Vert \bar{\bf Z}_{r_f} \Vert}^9} \cdot \sqrt{{\Vert \bar{\bf Z}_{r_f} \Vert}^2  {t^2_{go}} + 2  { \bar{\bf Z}^T_{r_f} \bar{\bf Z}_{V_f} }   {t_{go}} + {\Vert \bar{\bf Z}_{V_f} \Vert}^2}  
  \end{split}
\end{equation}
\begin{equation}
\begin{split}
        \bar{\bf I}_{B}(4:6)  
 &= \frac{-\alpha T_{\max}{\boldsymbol{ \mathcal K}}_2}{6  {\Vert \bar{\bf Z}_{r_f} \Vert}^7} \cdot \ln \left( \left|{\Vert \bar{\bf Z}_{r_f} \Vert}  \sqrt{{\Vert \bar{\bf Z}_{r_f} \Vert}^2  {t^2_{go}} + 2  { \bar{\bf Z}^T_{r_f} \bar{\bf Z}_{V_f} } {t_{go}} + {\Vert \bar{\bf Z}_{V_f} \Vert}^2} + {\Vert \bar{\bf Z}_{r_f} \Vert}^2  {t_{go}} + { \bar{\bf Z}^T_{r_f} \bar{\bf Z}_{V_f} } \right| \right ) 
\\
  &  + \frac{-\alpha T_{\max}{\boldsymbol{ \mathcal G}}_2(t_{go})}{6  {\Vert \bar{\bf Z}_{r_f} \Vert}^7} \cdot \sqrt{{\Vert \bar{\bf Z}_{r_f} \Vert}^2  {t^2_{go}} + 2  { \bar{\bf Z}^T_{r_f} \bar{\bf Z}_{V_f} }  {t_{go}} + {\Vert \bar{\bf Z}_{V_f} \Vert}^2}  
  \end{split}
\end{equation}
\end{subequations}
where
\begin{subequations}
\begin{equation}   
\begin{split}
    {\boldsymbol{ \mathcal K}}_1 & = \Bigg \{  
    {\bar{\bf Z}_{r_f}} \left[9  {\Vert {\bar{\bf Z}_{V_f}}  \Vert}^4  {\Vert \bar{\bf Z}_{r_f} \Vert}^4 - 90  \left({ \bar{\bf Z}^T_{r_f} \bar{\bf Z}_{V_f} } \right)^2  {\Vert \bar{\bf Z}_{V_f} \Vert}^2  {\Vert \bar{\bf Z}_{r_f} \Vert}^2 + 105  \left({ \bar{\bf Z}^T_{r_f} \bar{\bf Z}_{V_f} } \right)^4 \right]  
    \\
    & +  {\bar{\bf Z}_{V_f}} \left [ 36  { \bar{\bf Z}^T_{r_f} \bar{\bf Z}_{V_f} }  {\Vert \bar{\bf Z}_{V_f} \Vert}^2  {\Vert \bar{\bf Z}_{r_f} \Vert}^4 - 60  \left({ \bar{\bf Z}^T_{r_f} \bar{\bf Z}_{V_f} } \right)^3  {\Vert \bar{\bf Z}_{r_f} \Vert}^2 \right]  \Bigg \}    {a_2} \\ 
    & + \Bigg \{ { \bar{\bf Z}_{r_f}}  \left [ 36  { \bar{\bf Z}^T_{r_f} \bar{\bf Z}_{V_f} } {\Vert \bar{\bf Z}_{V_f} \Vert}^2  {\Vert \bar{\bf Z}_{r_f} \Vert}^4 - 60  \left({ \bar{\bf Z}^T_{r_f} \bar{\bf Z}_{V_f} } \right)^3  {\Vert \bar{\bf Z}_{r_f} \Vert}^2 \right ]  
    \\ &
    +{\bar{\bf Z}_{V_f}} \left [36  \left({ \bar{\bf Z}^T_{r_f} \bar{\bf Z}_{V_f} } \right)^2  {\Vert \bar{\bf Z}_{r_f} \Vert}^4 - 12  {\Vert \bar{\bf Z}_{V_f} \Vert}^2  {\Vert \bar{\bf Z}_{r_f} \Vert}^6 \right ]   \Bigg \}  {a_1} \\ 
    & + \left \{  {\bar{\bf Z}_{r_f}} \left [36  \left({ \bar{\bf Z}^T_{r_f} \bar{\bf Z}_{V_f} } \right)^2  {\Vert \bar{\bf Z}_{r_f} \Vert}^4 - 12  {\Vert \bar{\bf Z}_{V_f} \Vert}^2  {\Vert \bar{\bf Z}_{r_f} \Vert}^6 \right ]   - {\bar{\bf Z}_{V_f}}  \Bigg [ 24  { \bar{\bf Z}^T_{r_f} \bar{\bf Z}_{V_f} }   {\Vert \bar{\bf Z}_{r_f} \Vert}^6 \Bigg ] \right \}  {a_0} 
\end{split}
\end{equation}
\begin{equation}
   {{\boldsymbol{ \mathcal G}}_1}(t_{go}) =  {\boldsymbol \theta}_3(t_{go}) \cdot {t^3_{go}} + {\boldsymbol \theta}_2(t_{go}) \cdot {t^2_{go}} + {\boldsymbol \theta_1}(t_{go}) \cdot {t_{go}} + {\boldsymbol  \theta}_0
\end{equation}
\begin{equation}
    {\boldsymbol \theta_3}(t_{go}) = {\bar{\bf Z}_{r_f}}  \cdot  6  {a_2}  {\Vert \bar{\bf Z}_{r_f} \Vert}^7  
\end{equation}
\begin{equation}
    {\boldsymbol \theta_2} (t_{go}) = 
    \left [ {\bar{\bf Z}_{V_f}} \cdot  8  {\Vert \bar{\bf Z}_{r_f} \Vert}^7   - {\bar{\bf Z}_{r_f}} \cdot 14   {a_2}{ \bar{\bf Z}^T_{r_f} \bar{\bf Z}_{V_f} }   {\Vert \bar{\bf Z}_{r_f} \Vert}^5   \right] 
    + {\bar{\bf Z}_{r_f}}  \cdot 8  {a_1} {\Vert \bar{\bf Z}_{r_f} \Vert}^7 
\end{equation}
\begin{equation} 
    \begin{split}
    {{\boldsymbol \theta_1}(t_{go})}& = 
    \Bigg \{ {\bar{\bf Z}_{r_f}}  \Bigg [ 35  \left({ \bar{\bf Z}^T_{r_f} \bar{\bf Z}_{V_f} } \right)^2  {\Vert \bar{\bf Z}_{r_f} \Vert}^3 - 9  {\Vert\bar {\bf Z}_{V_f}  \Vert}^2  {\Vert \bar{\bf Z}_{r_f} \Vert}^5 \Bigg ] 
    - {\bar{\bf Z}_{V_f}} \cdot 20  { \bar{\bf Z}^T_{r_f} \bar{\bf Z}_{V_f} }  {\Vert \bar{\bf Z}_{r_f} \Vert}^5   \Bigg \}  {a_2} 
    \\&
    + \Bigg( {\bar{\bf Z}_{V_f}} \cdot  12  {\Vert \bar{\bf Z}_{r_f} \Vert}^7   - {\bar{\bf Z}_{r_f}} \cdot  20  { \bar{\bf Z}^T_{r_f} \bar{\bf Z}_{V_f} } {\Vert \bar{\bf Z}_{r_f} \Vert}^5   \Bigg)  {a_1} 
    + {\bar{\bf Z}_{r_f}} \cdot 12  {\Vert \bar{\bf Z}_{r_f} \Vert}^7    {a_0}   
    \end{split}
\end{equation}
\begin{equation} 
\begin{split}
   {\boldsymbol \theta_0} & = \Bigg \{ {\bar{\bf Z}_{r_f}} \Bigg [ 55  { \bar{\bf Z}^T_{r_f} \bar{\bf Z}_{V_f} } {\Vert \bar{\bf Z}_{V_f} \Vert}^2  {\Vert \bar{\bf Z}_{r_f} \Vert}^3 - 105  \left({ \bar{\bf Z}^T_{r_f} \bar{\bf Z}_{V_f} } \right)^3  {\Vert\bar {\bf Z}_{r_f}  \Vert} \Bigg]  
   \\ & 
   + {\bar{\bf Z}_{V_f}} \Bigg [ 60  \left({ \bar{\bf Z}^T_{r_f} \bar{\bf Z}_{V_f} } \right)^2  {\Vert \bar{\bf Z}_{r_f} \Vert}^3 - 16  {\Vert \bar{\bf Z}_{V_f} \Vert}^2  {\Vert \bar{\bf Z}_{r_f} \Vert}^5 \Bigg ]   \Bigg \}  {a_2} 
   \\
   &+ \Bigg \{ {\bar{\bf Z}_{r_f}} \Bigg [ 60  \left({ \bar{\bf Z}^T_{r_f} \bar{\bf Z}_{V_f} } \right)^2  {\Vert \bar {\bf Z}_{r_f}  \Vert}^3 - 16  {\Vert \bar {\bf Z}_{V_f}  \Vert}^2  {\Vert\bar {\bf Z}_{r_f}  \Vert}^5 \Bigg ]   - {\bar{\bf Z}_{V_f}} \cdot 36  { \bar{\bf Z}^T_{r_f} \bar{\bf Z}_{V_f} }  {\Vert \bar{\bf Z}_{r_f} \Vert}^5   \Bigg \}  {a_1} 
   \\&
   + \Bigg [{\bar{\bf Z}_{V_f}} \cdot  24  {\Vert \bar{\bf Z}_{r_f} \Vert}^7   -  {\bar{\bf Z}_{r_f}} \cdot  36  { \bar{\bf Z}^T_{r_f} \bar{\bf Z}_{V_f} }  {\Vert \bar{\bf Z}_{r_f} \Vert}^5  \Bigg ]  {a_0} 
   \end{split}
\end{equation}
\begin{equation}
\begin{split}
    {\boldsymbol{ \mathcal K}}_2 &= \Bigg \{ \bar{\bf Z}_{r_f}  \left [ 9   \bar{\bf Z}^T_{r_f}  \bar{\bf Z}_{V_f}  { \Vert \bar{\bf Z}_{V_f} \Vert}^2  { \Vert \bar{\bf Z}_{r_f} \Vert}^2 - 15  { \left( \bar{\bf Z}^T_{r_f}  \bar{\bf Z}_{V_f}  \right)}^3 \right ] \\ &
    + \bar{\bf Z}_{V_f} \left [ 9  { \left( \bar{\bf Z}^T_{r_f}  \bar{\bf Z}_{V_f}  \right)}^2  { \Vert \bar{\bf Z}_{r_f} \Vert}^2 - 3  { \Vert \bar{\bf Z}_{V_f} \Vert}^2  { \Vert \bar{\bf Z}_{r_f} \Vert}^4 \right ]   \Bigg \} a_2 
    \\& 
    +   \left \{  \bar{\bf Z}_{r_f}  \left [ 9  { \left( \bar{\bf Z}^T_{r_f}  \bar{\bf Z}_{V_f}  \right)}^2  { \Vert \bar{\bf Z}_{r_f} \Vert}^2 - 3  { \Vert \bar{\bf Z}_{V_f} \Vert}^2  { \Vert \bar{\bf Z}_{r_f} \Vert}^4 \right]  
    - \bar{\bf Z}_{V_f} \cdot  6   { \bar{\bf Z}^T_{r_f}  \bar{\bf Z}_{V_f}  }  { \Vert \bar{\bf Z}_{r_f} \Vert}^4   \right \} a_1
    \\ & 
    + \left(  \bar{\bf Z}_{V_f}  \cdot 6 { \Vert \bar{\bf Z}_{r_f} \Vert}^6 - \bar{\bf Z}_{r_f} \cdot 6 {\bar{\bf Z}^T_{r_f}  \bar{\bf Z}_{V_f}  }  { \Vert \bar{\bf Z}_{r_f} \Vert}^4  \right)  a_0 
\end{split}
\end{equation}
\begin{equation}
\begin{split}
    {\boldsymbol{ \mathcal G}}_2 (t_{go}) & = 
       \bar{\bf Z}_{r_f}\cdot  2 a_2 { \Vert \bar{\bf Z}_{r_f} \Vert}^5   \cdot   {t_{go}}^2 
      \\ & 
      + \left[ \left( \bar{\bf Z}_{V_f} \cdot 3 { \Vert \bar{\bf Z}_{r_f} \Vert}^5 -\bar{\bf Z}_{r_f} \cdot  5 \bar{\bf Z}^T_{r_f}  \bar{\bf Z}_{V_f} { \Vert \bar{\bf Z}_{r_f} \Vert}^3 \right) a_2 + \bar{\bf Z}_{r_f} \cdot 3a_1  { \Vert \bar{\bf Z}_{r_f} \Vert}^5   \right] \cdot  {t_{go}} 
      \\ &
      + \left \{ \bar{\bf Z}_{r_f}  \left [ 15{ \left( \bar{\bf Z}^T_{r_f}  \bar{\bf Z}_{V_f}  \right)}^2 { \Vert \bar{\bf Z}_{r_f} \Vert} - 4 { \Vert \bar{\bf Z}_{V_f} \Vert}^2  { \Vert \bar{\bf Z}_{r_f} \Vert}^3 \right]  -  \bar{\bf Z}_{V_f} \cdot 9  {  \bar{\bf Z}^T_{r_f}  \bar{\bf Z}_{V_f}  }  { \Vert \bar{\bf Z}_{r_f} \Vert}^3 \right \} \cdot  a_2 
      \\ & 
      +  \left ( \bar{\bf Z}_{V_f} \cdot 6  { \Vert \bar{\bf Z}_{r_f} \Vert}^5  -  \bar{\bf Z}_{r_f} \cdot  9 {\bf Z}^T_{r_f}   \bar{\bf Z}_{V_f}  { \Vert \bar{\bf Z}_{r_f} \Vert}^3  \right ) a_1 
      + \bar{\bf Z}_{r_f} \cdot 6 a_0 { \Vert \bar{\bf Z}_{r_f} \Vert}^5 
    \end{split}
\end{equation}
\end{subequations}
\emph{Remark:} The solutions of the integrals in \eqref{eqn:boundedintegralFO} are valid iff $\bar{\bf Z}_{r_f}$ and  $\bar{\bf Z}_{V_f}$ are not parallel.